\documentclass[fleqn,usenatbib]{mnras}

\usepackage{newtxtext,newtxmath}

\usepackage[T1]{fontenc}
\usepackage{ae,aecompl}

\usepackage{graphicx}	
\usepackage{amsmath}	
\usepackage{amssymb}	
\usepackage{newtxtext,newtxmath}
\usepackage[T1]{fontenc}
\usepackage[latin9]{inputenc}
\usepackage{color}
\usepackage{array}
\usepackage{booktabs}
\usepackage{units}
\usepackage{multirow}
\usepackage{amstext}
\usepackage{graphicx}
\usepackage[T1]{fontenc}
\usepackage[latin9]{inputenc}
\usepackage{color}
\usepackage{units}
\usepackage{textcomp}
\usepackage{amsmath}
\usepackage{amssymb}
\usepackage{graphicx}
\usepackage{savesym}
\usepackage{amsmath}
\usepackage{tabu}
\usepackage{longtable}
\usepackage{supertabular}
\usepackage{nicefrac}
\usepackage{pifont}
\usepackage[usenames,dvipsnames,svgnames,table]{xcolor}
\usepackage{pifont}
\usepackage{makecell}

\providecommand{\tabularnewline}{\\}
\savesymbol{iint}
\savesymbol{iiint}
\savesymbol{iiiint}
\let\oldAA\AA
\renewcommand{\AA}{\text{\normalfont\oldAA}}

\newcommand{\Xmark}{\ding{55}}
\defcitealias{Fernandez2018b}{F2018}

\title[A Bayesian direct method implementation]{A Bayesian direct method implementation to fit emission line spectra: Application to the primordial He abundance determination}

\author[V. Fern\'andez, E. Terlevich, A. I. D\'\i az, R. Terlevich]{Vital Fern\'andez$^{1}$\thanks{E-mail:
vi.fernandez@inaoep.mx (Vital Fern\'andez)}, Elena Terlevich$^{1}$, Angeles I. D\'\i az$^{2, 3, 4}$, 
Roberto Terlevich$^{1, 5}$\\
$^{1}$Instituto Nacional de Astrof\'\i sica, \'Optica y Electr\'onica, Luis E. Erro 1, 72840 Tonantzintla, Puebla, Mexico\\
$^{2}$Departamento de F\'isica Te\'orica, Universidad Aut\'onoma de Madrid, E-28049 Madrid, Spain\\
$^{3}$Centro de Investigaci\'on Avanzada en F\'isica Fundamental CIAFF-UAM \\
$^{4}$Astro-UAM, UAM, Unidad Asociada CSIC \\
$^{5}$ Institute of Astronomy, University of Cambridge, Madingley Rd., Cambridge CB3 0HA , UK}

\date{Accepted XXX. Received YYY; in original form ZZZ}

\pubyear{2015}

\begin{document}
\label{firstpage}
\pagerange{\pageref{firstpage}--\pageref{lastpage}}
\maketitle

\begin{abstract}
This work presents a Bayesian algorithm to fit the recombination and
collisionally excited line spectra of gas photoionized by clusters
of young stars. The current model consists in fourteen dimensions:
two electron temperatures, one electron density, the extinction coefficient,
the optical depth on the $HeI$ recombination lines and nine ionic
species. The results are in very good agreement with those previously
published using the traditional methodology. The probabilistic programming
library PyMC3 was chosen to explore the parameter space via a NUTs
sampler. These machine learning tools provided excellent convergence
quality and speed. The primordial helium abundance measured from a
multivariable regression using oxygen, nitrogen and sulfur was $Y_{P,\,O-N-S}=0.243\pm0.005$
in agreement with a standard Big Bang scenario.
\end{abstract}




\section{Introduction}

\begin{figure*}
\includegraphics[width=1\textwidth]{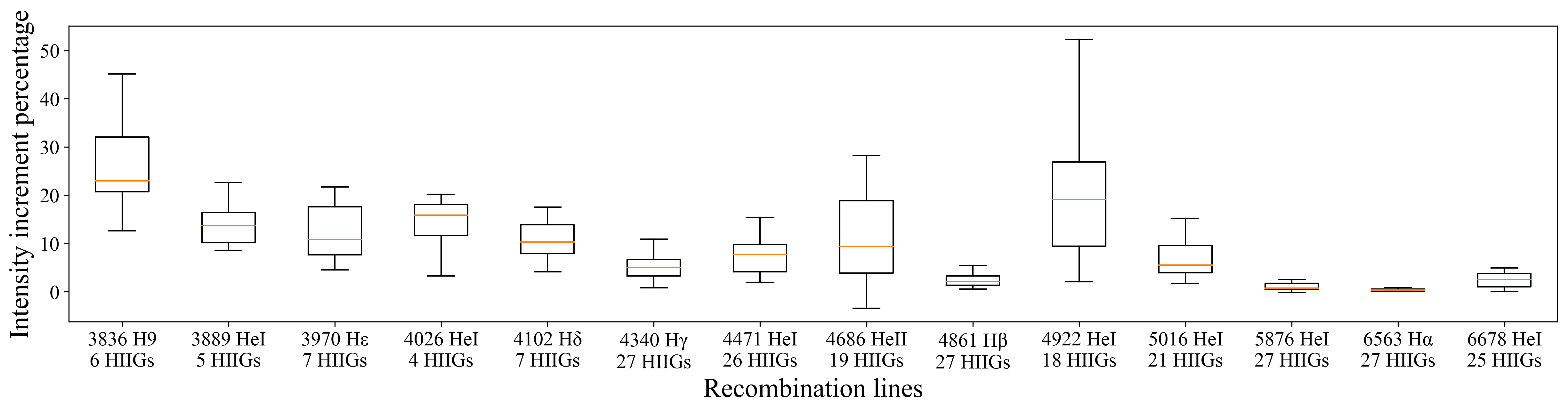}

\caption{\label{fig:Box-plot-with}Box and whisker plot with the percentage
increase in emission line flux after the continua correction has been
applied for some of the emission lines observed in increasing wavelength
and the number of HIIGs contributing to the statistic. The middle
line in the boxes represents the sample median increased percentage
while the box edges represent the $16^{th}$ and $84^{th}$ percentiles.
The whiskers represent the lower and upper outliers.}
\end{figure*}
The Standard Big Bang Nucleosynthesis (SBBN) is the model of choice
for explaining the non-zero helium mass fraction observed in the universe for objects with almost zero metals. Inside
its framework for a certain set of atomic parameters, the primordial
nucleosynthesis by-products $(D,\,^{4}He,\,{}^{3}He$ and $^{7}Li)$
can be modelled as a function of the baryons-to-photons ratio, $\eta$.
These are commonly referred to as light elements and an empirical
confirmation of their primordial abundances can provide very strong
constraints on the early universe. An indirect measurement for these
primordial abundances can be accomplished from the analysis of the
Cosmic Microwave Background (CMB) since $\eta$ can be measured from
its spectrum. Recently, the \citet{PlanckCollaboration2018} combined
their CMB measurements with the algorithms by \citet{Pisanti2008b}
to predict the light element abundances under a $\Lambda CDM$ paradigm.
Their prediction for the primordial helium mass fraction is $Y_{P}=0.24672_{-\left(0.00012\right)0.00061}^{+\left(0.00011\right)0.00061}$.
However, once the fitting is repeated allowing $Y_{P}$ and $N_{eff}$
(the number of neutrino species) to vary, the prediction uncertainty
increases to $Y_{P}^{BBN}=0.247_{-0.036}^{+0.034}$ with $N_{eff}=2.89_{-0.57}^{+0.63}$ using a neutron lifetime without uncertainties of $\tau_{n}=880.2 s$. 

The measurement of $Y_{P}$ involves the  analysis of
chemically unevolved gas reservoirs. In the recent work by \citet{Cooke2018e},
the helium abundance was measured from the absorption features of
intergalactic gas observed against the light of a background quasar. Their
$Y_{P}=0.250_{-0.025}^{+0.033}$ value is in agreement with the SBBN
\citet{PlanckCollaboration2018a} prediction. In the past, this technique
has provided very accurate measurements for the primordial deuterium
mass fraction \citep[see ][]{Carswell1994b} which may extend to the $Y_{P}$ estimation. Currently,
however, state of the art measurements for $Y_{P}$ are obtained from
the chemical analysis of the photo-ionised gas in extragalactic HII
regions. 

Over 40 years ago, \citet{Peimbert1974} and \citet{Lequeux1979}
published what is now known as the traditional method to determine
$Y_{P}$. These authors argued that in low metallicity regions the
chemical enrichment $\left(\nicefrac{dY}{dZ}\right)$ is linear. 
They proposed to use oxygen as the tracer for the total metal content
$\nicefrac{dY}{dZ}\approx\nicefrac{dY}{dO}$. This is a very convenient
choice for three reasons: firstly oxygen is the most common element
after hydrogen and helium, accounting for almost half the metals mass
fraction, $Z$. Secondly, in the optical range oxygen has very intense
lines, which provide an accurate measurement for its main ionic populations
$\left(O^{+},\,O^{2+}\right)$. Finally, oxygen is a heavy element
produced by very massive stars in very short time scales. This guarantees
that the gas reservoir is not largely contaminated by the current
star-forming burst. These assumptions make it possible to apply a
linear regression on the $Y$ vs $\nicefrac{O}{H}$ abundances relation
for a given data sample. The coordinate $\nicefrac{O}{H}=0$ corresponds
to $Y=Y_{P}$: the helium abundance produced by the primordial nucleosynthesis.

Recent primordial helium abundance determinations using this technique
have been published by \citet{Izotov2014b} with $Y_{P}=0.2551\pm0.0022$,
\citet{Aver2015b} with $Y_{P}=0.2449\pm0.004$, \citet{Peimbert2016b}
with $Y_{P}=0.2446\pm0.0029$ and \citet{Valerdi2019} with $Y_{P}=0.2451\pm0.0026$. The last works from Peimbert and collaborators include a computation of the neutron lifetime using the measured value $Y_{P}$ and a SBBN model. The estimated values: $\tau_{n}=870\pm14$ and  $\tau_{n}=873\pm14$  are in good agreement with the value measured by \cite{Pattie2018}  $\tau_{n}=877.7\pm0.7_{stat}+0.4/-0.2_{sys}$.
Additionally, \citet{Fernandez2018b} (hereafter \citetalias{Fernandez2018b})
published the first $Y_{P}$ determination using sulfur as the metallicity
tracer resulting in $Y_{P,S}=0.244\pm0.006$. A multivariable
linear regression using oxygen, nitrogen and sulfur was proposed giving
$Y_{P,\,O-N-S}=0.245\pm0.007$. It can be appreciated that the latter
three groups measurements agree with a SBBN scenario, while the value
from \citet{Izotov2014b} is $2.2\sigma$ above the standard cosmology
predictions. This discrepancy may be explained by the complexity in
the chemical analysis. \citet{Peimbert2011} summarised thirteen uncertainty
sources, most of them systematic, which affect the helium abundance
calculation. Many of these error sources also impact the
metals abundance determination. These include the uncertainty on the
atomic data, the temperature and density distributions or the accuracy
on the ionisation correction factors.

The methodology described in \citetalias{Fernandez2018b} follows
a standard approach to derive the chemical composition of ionised
gas. This procedure, recently reviewed by \citet{Peimbert2017b} and
\citet{Perez-Montero2017b} is designed to isolate each phenomenon
contributing to the observed emission. However, as both the astronomical
data and models complexity keep increasing, this methodology becomes
more cumbersome to apply. This is the reason why many researchers
are embracing methodologies capable of exploring a multidimensional
space. Examples of chemical analysis in the optical spectral range
include the pioneer work by \citet{Olive2001b} and \citet{Olive2004b}
which fitted simultaneously five parameters contributing to the helium
lines emission. The aim of this new methodology is to provide a better
quantification of the errors for the helium abundance determination.
Later, this methodology was further enhanced by \citet{Aver2011a,Aver2013,Aver2015b},
who included the hydrogen emission lines and increased the number
of dimensions to eight. They also discussed the impact of the computational
technique in order to explore the parameter space. Their chosen sampler
was COSMOMC \citep[see][]{Li2014b}, a Fortran based program which
applies a Markov Chain Monte Carlo (MCMC) process via a Metropolis-Hastings
algorithm.

In this paper, we propose a methodology to fit the recombination and
collisionally excited emission spectra. As it was done in \citetalias{Fernandez2018b},
the electron temperature and density are determined using the direct
method \citep[see][]{Osterbrock2005} which are anchored by appropriate
collisionally excited line ratios. The main advantage of the new method
is that all the model parameters contributing to the observed emission
are solved simultaneously. This represents a space with up to fourteen
dimensions. These parameters include nine ionised species of argon,
helium, oxygen, nitrogen and sulfur in the optical and near infrared
regions. To solve this system, we apply a HMC (Hamiltonian Monte Carlo)
sampler based in machine learning algorithms. These tools provide
a remarkable convergence quality and speed, which would have been
impossible with standard MCMC algorithms. The data used is the one
described in \citetalias{Fernandez2018b}.

The paper is organised as follows: Section 2 briefly reviews the \citetalias{Fernandez2018b}
sample and the treatment to account for the stellar and nebular continua.
Section 3 describes how the direct method was adapted to a Bayesian
paradigm and a technical description of the HMC sampler. Section 4
overviews a set of test cases to evaluate the accuracy and physical
viability of the new technique. Section 5 presents the results and
discusses merits and limitations of this technique. Finally, Section
6 summarises the main results.

\section{The data}\label{The-data}

The data used in this paper are the 27 HII galaxies presented in Table
1 of \citetalias{Fernandez2018b}. In that paper we discussed the
selection of the objects, the data reduction process and the derivation
of their chemical abundances. The trends $Y-O/H$, $Y-N/H$ and $Y-S/H$
and the obtained values for $Y_{P}$ are also discussed in that paper.

The measurement of the line fluxes has been described in detail in
\citetalias{Fernandez2018b}. The process takes into account corrections
for extinction and for underlying stellar and nebular continua. For
the starforming ionised regions, the nebular continuum can reach comparable
values to the underlying stellar continuum \citep[see][]{Reines2009}.

The impact of this treatment of the spectra is illustrated in Fig.
\ref{fig:Box-plot-with}, where we plot the percentage increase in
the line fluxes after the corrections have been applied for the most
relevant recombination lines in our spectra. Even though the contribution
for the strong hydrogen Balmer lines may be considered negligible,
we observe about a 20 \% increase in the HeI lines intensity once
the continuum correction is taken into account and this percentage
changes with wavelength.

\section{Bayesian direct method.}\label{Bayesian-direct-method}

The flux from an emission line can be expressed as a function of the
transition emissivity relative to a Balmer line, e.g. $H\beta$:
\begin{equation}
\frac{F_{X^{i+},\,\lambda}}{F_{H\beta}}=X^{i+}\frac{\epsilon_{X^{i+},\,\lambda}\left(T_{e},\,n_{e}\right)}{\epsilon_{H\beta}\left(T_{e},\,n_{e}\right)}\cdot10^{-c\left(H\beta\right)\cdot f_{\lambda}}\cdot k_{HeI,\,\lambda}\left(\tau,\,T_{e},\,n_{e}\right)\label{eq:fluxFormula}
\end{equation}
where $\nicefrac{\epsilon_{X^{i+},\,\lambda}}{\epsilon_{H\beta}}$
is the relative emissivity at the transition wavelength $\lambda$,
for an ion with abundance $X^{i+}$, at certain electron temperature
$T_{e}\,\left(K\right)$ and electron density $n_{e}$~$\left(cm^{-3}\right)$.
The term $c\left(H\beta\right)$ is the logarithmic extinction at
$H\beta$ for a reddening law $f_{\lambda}$. The parameter $k_{HeI,\,\lambda}$
represents a fluorescence excitation correction for the $HeI$ lines
calculated by \citet{Benjamin2002b} as a function of the optical
depth $\tau$.

Traditionally, as already mentioned, the parameters in eq. \ref{eq:fluxFormula}
are solved individually by isolating the different processes contributing
to the observed flux. Usually, the first step consists in computing
$c\left(H\beta\right)$ from the difference between the theoretical
and observed fluxes of the hydrogen Balmer series. This is a valid
approach since most of the hydrogen is ionised. Additionally, their
emissivity remains almost constant for the expected $T_{e}$ and $n_{e}$
values. 

The next step in the traditional method consists in computing the
ionic abundances from the observed emission lines. In the so called
direct method, we calculate the emissivity coefficient by providing
values for $T_{e}$ and $n_{e}$ obtained by measuring temperature-sensitive
and density-sensitive line ratios. Finally, using these parameters
back in eq. \ref{eq:fluxFormula} the ionic abundance can be computed
from each emission line.

As we discussed in \citetalias{Fernandez2018b}, however, there are
problems inherent to this method. Firstly, it can underestimate the
uncertainty in the model parameters since these are fitted individually.
To properly propagate the error to the abundances, several iterations
may be necessary, as well as complex networks of chained Monte Carlo
algorithms. Secondly, eq. \ref{eq:fluxFormula} represents the most
basic interpretation. As we introduce more complexity in the physics
of the problem, like e.g. $k_{\tau}$ in eq. \ref{eq:fluxFormula},
it becomes harder to solve the model parameters. Finally, with the
large amount of data available today, plus its larger wavelength coverage,
this methodology becomes harder to apply. 

In this paper, we use a method that allows to fit all the parameters
in eq. \ref{eq:fluxFormula} simultaneously. These are two electron
temperatures for the low and high ionisation regions, $T_{low}$ and
$T_{high}$, one electron density $n_{e}$, which is assumed to remain
constant within the star forming region, the extinction coefficient
$c\left(H\beta\right)$ and nine ionic abundances $Ar^{2+}$, $Ar^{3+}$,
$y^{+}$, $y^{2+}$, $O^{+}$, $O^{2+}$, $N^{+}$, $S^{+}$ and $S^{2+}$.
Here we add the optical depth, $\tau$, to account for the fluorescence
excitation in the helium lines as tabulated by \citet{Benjamin2002b}.
This brings the maximum number of dimensions to fourteen and in order
to explore this relatively large parameter space, it becomes necessary
to apply machine learning tools. 

\begin{table}
\caption{\label{tab:Priors-and-likelihood}Priors and likelihood distributions
in our model. The term $X^{i+}$ includes all the ionic metal abundances:
$Ar^{2+}$, $Ar^{3+}$, $y^{+}$, $y^{2+}$, $O^{+}$, $O^{2+}$,
$N^{+}$, $S^{+}$ and $S^{2+}$. These abundances are define in $12+log\left(X^{i+}\right)$
scale.}

\centering{}%
\begin{tabular}{cc}
\toprule 
Parameter & Prior distribution\tabularnewline
\midrule
$T_{low}$ & $Normal(\mu=15000\,K,\,\sigma=5000\,K)$\tabularnewline
$T_{high}$ & $Normal(\mu=15000\,K,\,\sigma=5000\,K)$\tabularnewline
$n_{e}$ & $Normal\left(\mu=150\,cm^{-3},\,\sigma=50\,cm^{-3}\right)$\tabularnewline
$c(H\beta)$ & $logNormal\left(\mu=0,\,\sigma=1\right)$\tabularnewline
$X^{i+}$ & $Normal(\mu=5,\,\sigma=5)$\tabularnewline
$y^{+}$ & $\nicefrac{1}{10}\cdot logNormal\left(\mu=0,\,\sigma=1\right)$\tabularnewline
$y^{2+}$ & $\nicefrac{1}{1000}\cdot logNormal\left(\mu=0,\,\sigma=1\right)$\tabularnewline
$\tau$ & $logNormal\left(\mu=0,\,\sigma=0.4\right)$\tabularnewline
\midrule 
Parameter & Likelihood distribution\tabularnewline
\midrule
$\frac{F_{X^{i+},\,\lambda}}{F_{H\beta}}$ & $Normal(\mu=\frac{F_{X^{i+},\,\lambda,\,obs}}{F_{H\beta}},\sigma=\frac{\sigma_{X^{i+},\,\lambda,\,obs}}{F_{H\beta}})$\tabularnewline
\bottomrule
\end{tabular}
\end{table}
\begin{table*}
\caption{\label{tab:testCasesResults} Fitting results for the test cases with
increasing number of parameters described in the text }

\centering{}%
\begin{tabular}{cccccc}
\hline 
Parameter & True value & Test 1 & Test 2 & Test 3 & Test 4\tabularnewline
\hline 
$T_{low}$ & 15590 & $15590\pm244$ & $15750\pm252$ & $16470\pm787$ & $15640\pm282$\tabularnewline
$n_{e}$ & 500 & $491\pm39$ & $492\pm39$ & $494\pm39$ & $492\pm36$\tabularnewline
$S^{+}$ & 5.48 & $5.479\pm0.014$ & $5.459\pm0.028$ & $5.402\pm0.064$ & $5.475\pm0.019$\tabularnewline
$S^{2+}$ & 6.36 & $6.360\pm0.015$ & $6.332\pm0.036$ & $6.258\pm0.084$ & $6.354\pm0.022$\tabularnewline
$O^{+}$ & 7.80 & \Xmark & $7.768\pm0.047$ & $7.663\pm0.116$ & $7.794\pm0.037$\tabularnewline
$O^{2+}$ & 8.05 & \Xmark & $8.037\pm0.022$ & $8.027\pm0.026$ & $8.048\pm0.017$\tabularnewline
$Ar^{2+}$ & 5.72 & \Xmark & $5.696\pm0.031$ & $5.634\pm0.071$ & $5.715\pm0.019$\tabularnewline
$Ar^{3+}$ & 5.06 & \Xmark & $5.049\pm0.019$ & $5.043\pm0.021$ & $5.058\pm0.016$\tabularnewline
$N^{+}$ & 5.84 & \Xmark & $5.820\pm0.027$ & $5.762\pm0.064$ & $5.835\pm0.019$\tabularnewline
$c(H\beta)$ & 0.100 & \Xmark & $0.137\pm0.047$ & $0.21\pm0.09$ & $0.11\pm0.018$\tabularnewline
$T_{high}$ & 16000 & \Xmark & \Xmark & $16310\pm334$ & $16020\pm193$\tabularnewline
$y^{+}$ & 0.0850 & \Xmark & \Xmark & \Xmark & $0.0850\pm0.001$\tabularnewline
$y^{2+}$ & 0.00088 & \Xmark & \Xmark & \Xmark & $0.00088\pm0.00001$\tabularnewline
$\tau$ & 1.0 & \Xmark & \Xmark & \Xmark & $0.991\pm0.225$\tabularnewline
\hline 
\end{tabular}
\end{table*}
Currently, data science has become very popular due to the computational
advances in three domains: big data, deep learning and probabilistic
programming. In the first field, the user is interested in finding
patterns in large data sources. In deep learning, the user defines
the parameter space via neural networks which have several `depths'.
Finally, probabilistic programming focuses in declaring models, whose
parameters and outputs are probability functions. We applied the probabilistic
programming package PyMC3 by \citet{Salvatier2016b}, which makes
use of the deep learning library Theano by \citet{TheTheanoDevelopmentTeam2016b}
to define the physical model. This package includes a NUTs (No-U-Turns)
sampler \citep[see][]{Hoffman2011b} to explore the parameter space.
This sampler follows a Hamiltonian Monte Carlo (HMC) paradigm, which
drops the stochastic jumps from the Markov Chain by an informed sampling
guided by the mathematical model derivatives. This algorithm implementation
involves a more challenging programming experience than a MCMC sampler.
Still, this HMC algorithm provides excellent convergence quality for
large parameter spaces. Moreover, this sampler decreases the fitting
computational time from several hours $\left(>6\,h\right)$ to a couple
of minutes.

It may be inferred from the previous definition that any probabilistic
programming implementation is intrinsically Bayesian. This paradigm
is characterised by the application of the Bayesian theorem:
\begin{equation}
Pr\left(\theta|y\right)=\frac{Pr\left(y|\theta\right)Pr\left(\theta\right)}{Pr\left(y\right)}\label{eq:bayesTheorem}
\end{equation}
where $Pr\left(\theta\right)$ is referred to as the prior. This term
represents the probability distribution of the model parameters $\theta$
before the fitting. $Pr\left(y|\theta\right)$ is the likelihood of
the model and it provides an evaluation on how the observational data
$y$ adjusts to the theoretical model. $Pr\left(y\right)$ is referred
to as the evidence. This parameter is actually the integral of the
numerator in eq. \ref{eq:bayesTheorem} over the complete parameter
space and it represents the probability that the observed data has
been generated by the processes described by the model. Finally, $P\left(\theta|y\right)$
is the probability of the model parameters given the observations.
In a Bayesian inference, the posterior provides the user with a credible
region: given the observed data, there is a $95\%$ probability that
the true value of $\theta$ falls within this credible region $CR_{\theta}$.
To successfully compute this diagnostic, however, it is essential
to properly define the terms in the Bayesian theorem.

Recently, \citet{Tak2018b} reviewed the quality of Bayesian models
in astronomical literature. These authors emphasised how improper
priors can result in output distributions not meeting the posterior
propriety. An example of improper prior includes a uniform distribution
covering the real space $\left(-\infty,\,\infty\right)$. Moreover,
in the particular case of a NUTs sampler a uniform prior can dramatically
affect both the convergence quality and the simulation speed. To deal
with this issue, in the Bayesian approach, one can define a joint
distribution using proper priors to ensure the posterior propriety.
On the one hand, we can rely on the scientific evidence to implement
very informative (or constrained) prior distributions. On the other
hand, in cases where little physical knowledge is available, it is
acceptable to provide uninformative (or wide) prior distributions.
In practice, this provides a uniform probability distribution for
a region of interest. These two strategies are considered here and
the input priors are displayed in Table \ref{tab:Priors-and-likelihood}.
The following paragraphs describe how the traditional direct method
applied in \citetalias{Fernandez2018b} was adjusted to a Bayesian
paradigm using proper priors:
\begin{itemize}
\item We consider in this study two ionisation regions. These are characterised
by two electron temperatures: $T_{low}$ and $T_{high}$. In the low
ionisation region the ionised species are $Ar^{2+}$, $H^{+},$ $O^{+}$,
$N^{+}$, $S^{+}$ and $S^{2+}$. The high ionisation species are
$Ar^{3+}$, $He^{+}$, $He^{2+}$ and $O^{2+}$. The two temperatures
share the same prior design: a Gaussian distribution with $\mu=15000\,K$
and $\sigma=5000\,K$. This range provides a good coverage for the
temperatures commonly encountered in HII galaxies (as it was shown
in \citetalias{Fernandez2018b}). The temperature depends on the auroral
lines observed: $\left[SIII\right]6312\AA$ for $T_{low}$ and
$\left[OIII\right]4363\AA$ for $T_{high}$. In the objects where
only one auroral line was available, only the corresponding temperature
priors is declared. The other temperature is calculated using the
linear relation provided by \citet{Perez-Montero2017b}:
\end{itemize}
\begin{equation}
t_{e}\left[OIII\right]=1.0807t_{e}\left[SIII\right]-0.0846\label{eq:TOIII-TSIII-relation}
\end{equation} where $t_{e}$ is in $10^{4}K$ units.

\begin{itemize}
\item The electron density calculation from the $\left[SII\right]\lambda\lambda6717$,$6731\AA$
line ratio falls within the very low density regime for most objects
$\left(<\,100\,cm^{-3}\right)$. This is actually the regime at which
this line ratio becomes less sensitive to density. This means that
the $\left[SII\right]$ density cannot  be computed to high accuracy.
This is not, however, a big issue from the chemical analysis point
of view as for most ionised species the emissivity also remains independent
with the density at these regimes. This can be an issue from the mathematical
point of view, though. If the density cannot be fitted from the available
data, the resulting posterior distribution will actually be the prior
one. In order to force the simulation to stay in a representative
density region, we define the density prior via a Gaussian distribution
with $\mu=n_{\left[SII\right]}$ and $\sigma=\sigma_{n_{e}\left[SII\right]}$
measured in \citetalias{Fernandez2018b}.
\item The reddening curve chosen for this work is the 'LMC average' published
by \citet{Gordon2003b} with a $R_{V}=3.4$ appropriate for a star
forming region. The gas extinction in these objects is usually very
low \citep[e.g.][]{Terlevich1991b}. This parameter, however, cannot
be negative. To account for this physical limit a log-normal distribution
was chosen with $\sigma=1.0$. This guarantees higher probability
to the expected extinction coefficient range.
\item In \citetalias{Fernandez2018b}, all calculations involving the line
emissivities were accomplished using PyNeb \citep[see][]{Luridiana2015b}.
Third party libraries, however, are not easily imported into deep
learning algorithms. Implementing a 2-dimensional interpolation
on emissivity grids is beyond the scope of this project. The alternative
chosen is a strategy commonly used in the traditional chemical analysis:
Parametrised equations for the emissivity as a function of the electron
temperature and density. For each emission line an emissivity grid
was computed using PyNeb for the references in Table \ref{tab:Parametrised-relations-and}
in the appendix and which also includes the emissivity
parametrisation for each ion.  Table \ref{tab:Parametrised-coefficients}
displays the fitted coefficients for the $8,000$ - $25,000\,K$ and
$1$-$600\,cm^{-3}$ surface for each emission line. These fittings
provide the line emissivity in log scale. This was done for two reasons:
first, the emissivity grids are easier to fit in a log scale surface.
Second, to improve the metal abundance sampling it was desirable to
use the standard $12+log\left(X^{+}\right)$ notation. For the ionic
abundances uninformative priors can be applied. For the metal species
Gaussian distributions are considered with $\mu=5$ and $\sigma=5$.
The $y^{+}$ and $y^{2+}$ abundances are given in the linear scale
and a log-normal distribution is used to model their priors with $\mu=0$
and $\sigma=1$. The values drawn from these priors are scaled by
the expected $y^{+}$ and $y^{2+}$ regime via coefficients $k_{y^{+}}=\nicefrac{1}{10}$
and $k_{y^{2+}}=\nicefrac{1}{1000}$.
\item 
\begin{figure}
\includegraphics[width=1\columnwidth]{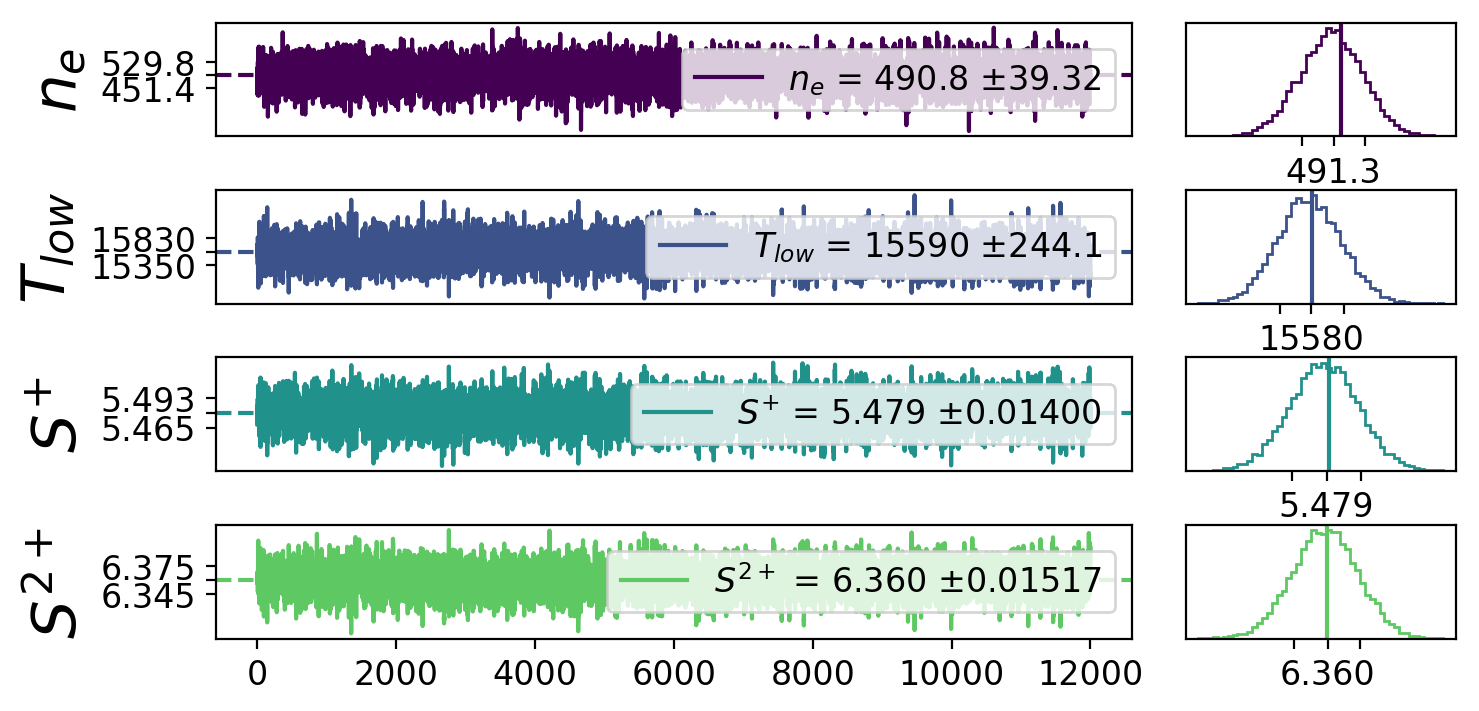}

\caption{\label{fig:test-case1}Output fit plot for the $1^{st}$ test case:
sulfur emission only. In colour in the electronic version.}
\end{figure}
To account for  fluorescence excitation on the helium lines, a
new correction has been included on the analysis. The radiative transfer
calculations from \citet{Benjamin2002b} are included in eq. \ref{eq:fluxFormula}
as a function of $T_{high}$ and $n_{e}$. A successful fitting for
this parameter depends on the availability of $HeI$ lines which are
sensitive to this effect. 
As it will be showed in the next section test cases, this parameter fits properly in synthetic observations for both prior probability distributions considered. However, our current observational data lacks emission lines which are strongly affected by this phenomenon. Therefore, it cannot be quantified. Consequently, we defined a prior with a log-normal distribution with $\mu=0$ and $\sigma=0.4$. This choice is justified by the following arguments: 1) In \citet{Aver2012a}, it was shown that more than half of the fifty successfully fitted spectra using the same optical depth model have an opacity below one and only three objects displayed opacities above four. Therefore, this distribution guarantees a proper prior, even if very informative, which is justified by known data. This is an acceptable practice for statistical inference in scientific models. 2) Using synthetic test cases, it was confirmed that this probability distribution does not result in degeneracies with other parameters, in particular $y^{+}$. This was not the case of prior designs, which covered large $\tau$ values.
\end{itemize}
The last element to define in eq. \ref{eq:bayesTheorem} is the model
likelihood. At this point, it is important to remember that in the
Bayesian paradigm there is not uncertainty in the model data $y$:
any randomness in the observables is caused by the model parameters,
which behave as a probability distribution. This interpretation has
a physical foundation in models such as this one. For example: $T_{e}$
is generally interpreted as the result of a Maxwellian velocity distribution
of the electrons. In practice, it is not common for spectra to show
the same uncertainty everywhere along the wavelength range. For example,
in the present case it is essential to assert the difference in uncertainty
between the $\left[OIII\right]$ auroral and nebular lines. A valid
approach to account for this uncertainty consists in establishing
a normal distribution, whose standard deviation is weighted by the
observational error, as the likelihood. This was the approach considered
in this model (shown in Table \ref{tab:Priors-and-likelihood}) with
a normal distribution likelihood with $\mu=F_{\lambda}$
and $\sigma=\sigma_{F_{\lambda}}$ for each emission line where  $F_{\lambda}$ represents the flux relative to $H\beta$ including the error propagation in the ratio.

\section{Model test cases}\label{Model-test-cases}

\begin{figure*}
\includegraphics[width=1\columnwidth]{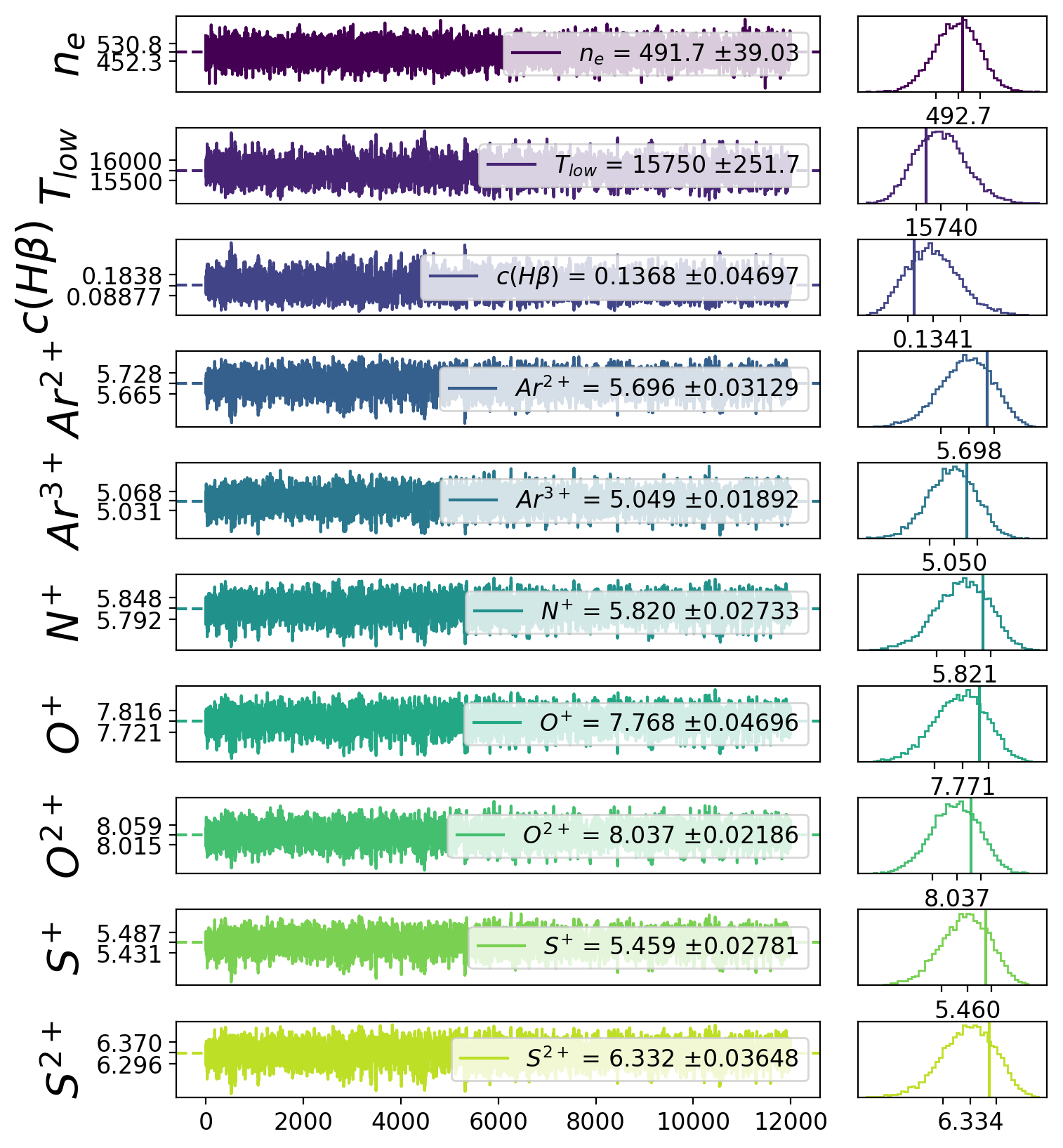}\includegraphics[width=1\columnwidth]{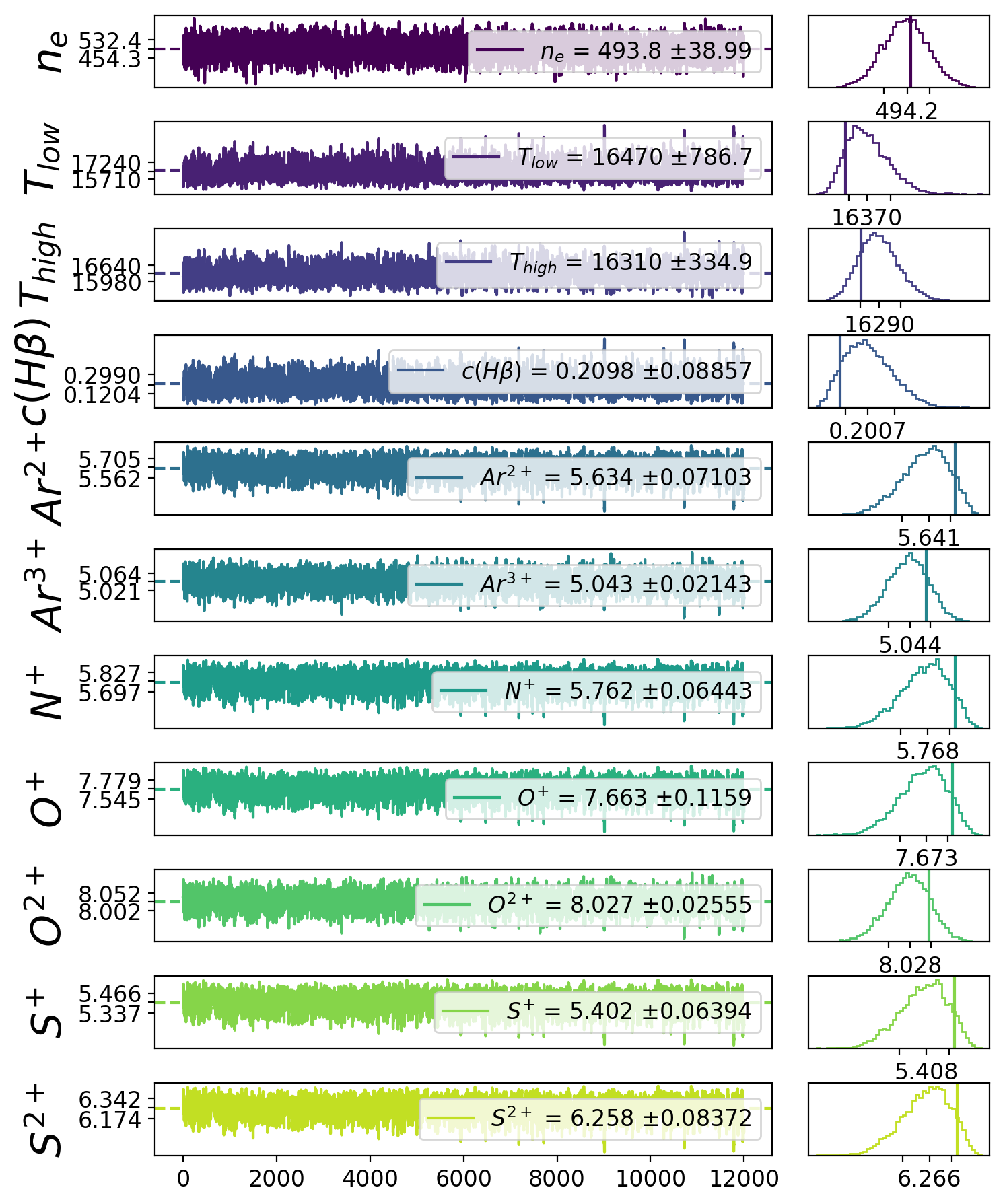}

\caption{\label{fig:test-case2-3} Left) Output fit plot for the $2^{nd}$
test case: all metals emission and light extinction. Right) Output
fit plot for the $3^{rd}$ test case: two electron temperatures. In colour in the electronic version.}
\end{figure*}
\begin{figure}
\includegraphics[width=1\columnwidth]{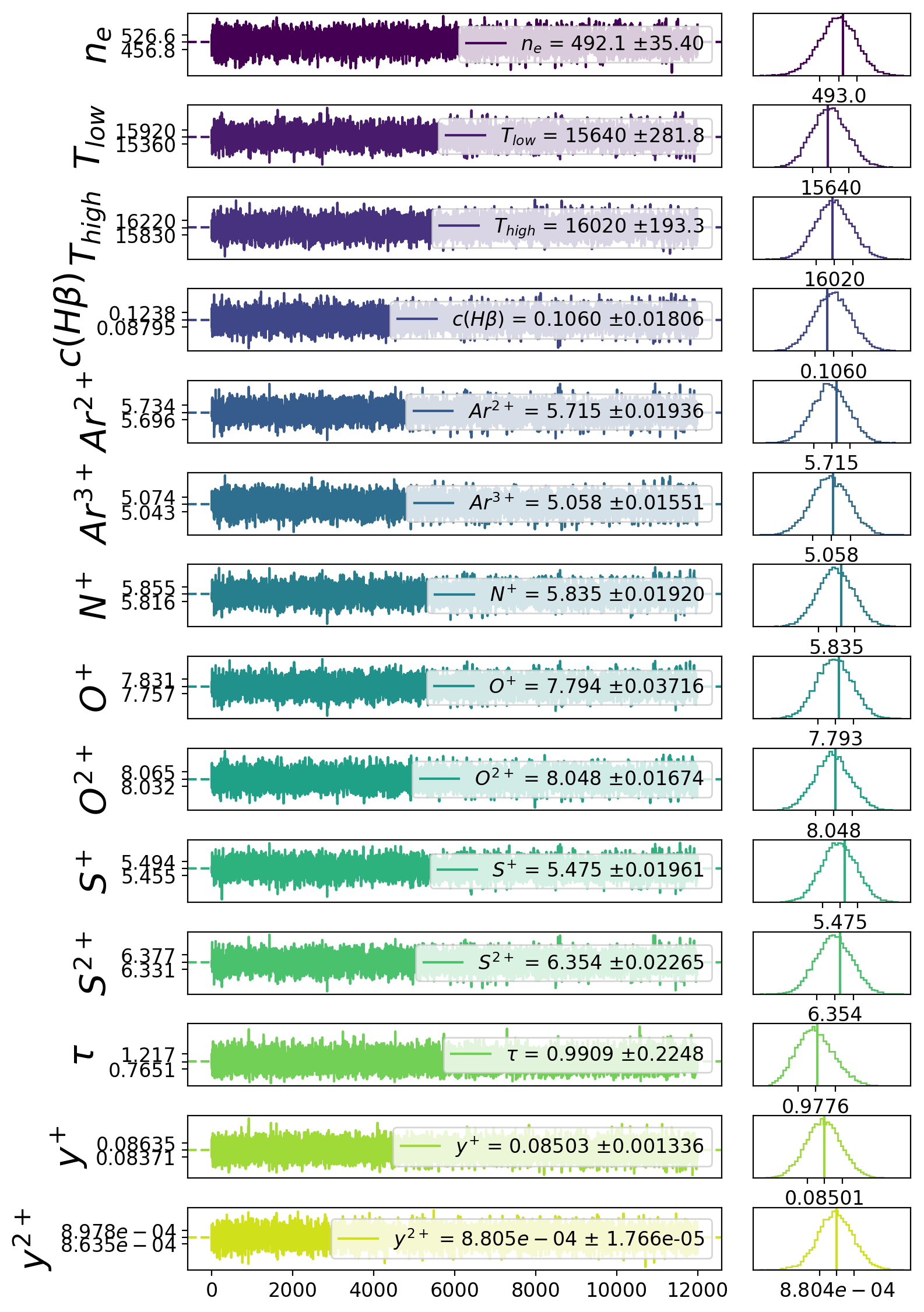}

\caption{\label{fig:test-case4}Output fit plot for the $4^{th}$ test case:
complete model. In colour in the electronic version.}
\end{figure}
\begin{figure*}
\includegraphics[width=1\textwidth]{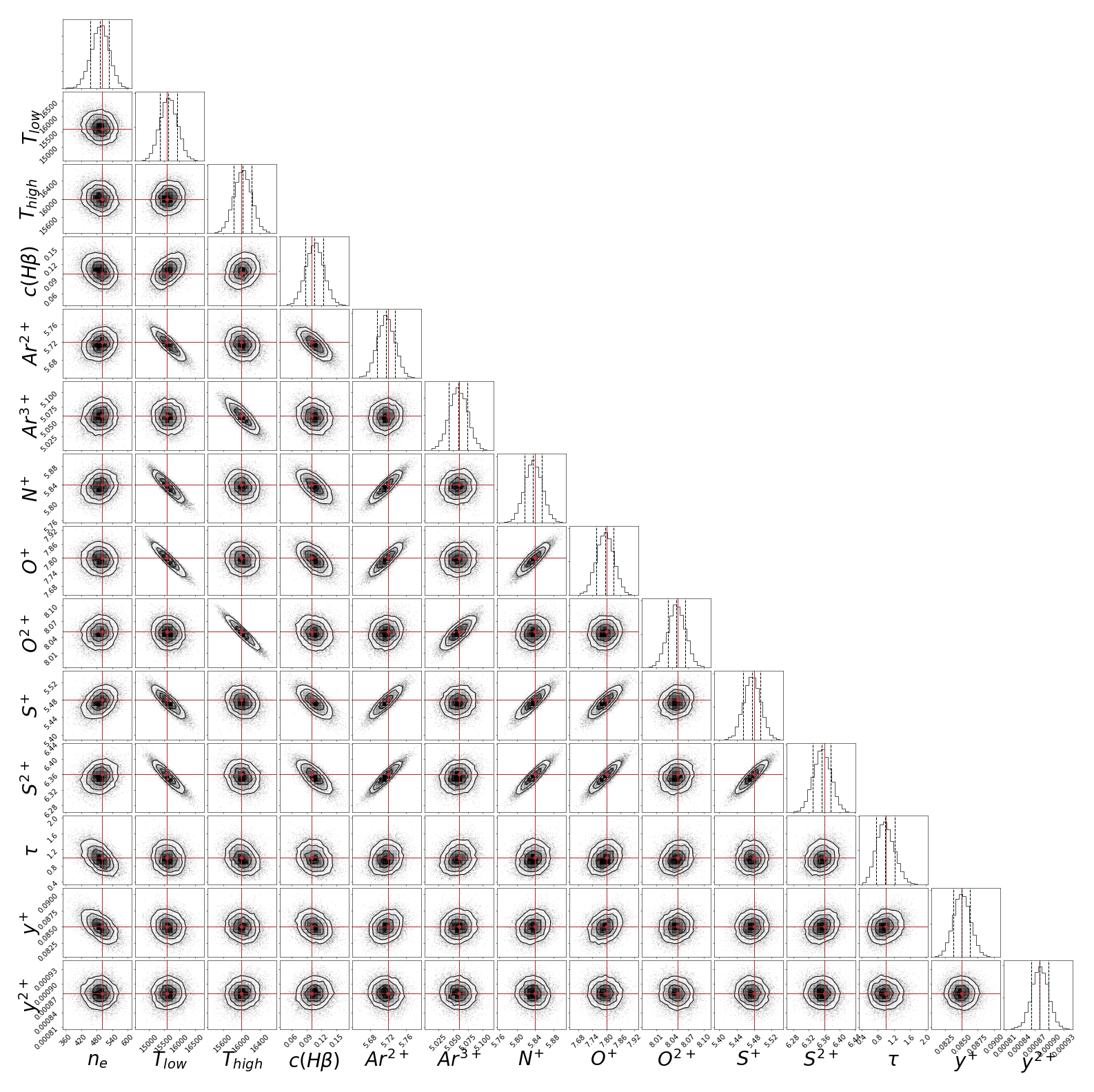}

\caption{\label{fig:corner_plot_test-case4}Scatter plot matrix for the $4^{th}$
test case with a complete model. The red crosses over the surface
plots represent the true values used to compute the synthetic fluxes.
This plot has been created using the corner library by \citet{Foreman-Mackey2016}.}
\end{figure*}
Possibly the most discouraging feature of Monte Carlo simulations
is assessing the convergence quality. Fortunately, in empirical models
such as this one, the convergence can be evaluated graphically. This
process has two phases:

In the first one, a set of values is chosen for the model parameters
from which the synthetic observables are generated. Afterwards, this
synthetic spectrum is fed into the algorithm: The output fittings
are then compared with the true values. In addition to a good accuracy
it is expected that the credible region width is of the same order
as the uncertainty found in the traditional methodology. Table \ref{tab:testCasesResults}
displays in column 1 a set of values for the model parameters, which
were used to generate an array of synthetic emission lines using eq.
\ref{eq:fluxFormula}. In the test cases below, a uniform $2\%$ uncertainty
was adopted for the emission lines flux considered. The following
paragraphs describe these test cases.

In the simplest example only four model parameters are fitted $\left(T_{low},\,n_{e},\,S^{+},S^{2+}\right)$
using five emission lines: $\left[SII\right]\lambda\lambda6717$,$6731\AA$,
$\left[SIII\right]\lambda6312\AA$ and $\left[SIII\right]\lambda\lambda9069,9531\AA$.
In a Monte Carlo simulation, such as this one, the user declares the
number of iterations and the number of tuning steps, which are the
number of initial steps not included in the final statistics. This
is because they account for the time the algorithm requires to reach
the optimum jump step size to the solution region. An additional good
practice consists in repeating the simulation to confirm that the
initial conditions do not alter the results. Fortunately, packages
such as PyMC3 can parallelize the models easily by running one simulation
per machine processor. Once all the fittings finish, the parameters
measurement includes the traces from all the machine cores. This enables
shorter simulations and a better resource management. All this information
can be seen in Fig.\ref{fig:test-case1}. The four plots on the left
side represent the model parameter traces evolution with time. It
can be appreciated that during the simulation steps, all the traces
have a white noise-like appearance. This signals a simulation which
has successfully converged to a parameter space coordinate (not necessarily
the right one). This simulation configuration consists in 6000 simulation
steps and 2000 tuning steps parallelised in a 1.8-GHz i7-5400U dual-core processor laptop.
These simulation pairs are joined in Fig.\ref{fig:test-case1} where
we can see that the traces are indistinguishable. These traces provide
us with the statistics for the parameters measurement. The value written
over the traces are the mean and standard deviation from the traces.
These are the values in Table \ref{tab:testCasesResults}. The $16^{th}$
and $84^{th}$ percentiles from the posterior distribution are written
on the ordinate axis. The median value is given in the histograms.
It can be concluded from this initial test that the simulation fittings
are in good agreement with the true values.

Fig. \ref{fig:test-case2-3} displays the results for the test cases
2 and 3 similarly to Fig.\ref{fig:test-case1}, but now all the metal
emission lines are included as input. Consequently, all metals ionic
abundances are included as well. In this simulation, we still only have one temperature prior  $(T_{low})$. In order to compute the physically correct temperature for the high ionization ions, the algorithm automatically includes the empirical linear model in eq. \ref{eq:TOIII-TSIII-relation}.  Hence, at each iteration the temperature value drawn from this distribution is applied directly to compute the emissivity for the low ionization species emissivity. For the high ionization ones, however, this temperature is previously corrected according to eq. \ref{eq:TOIII-TSIII-relation}.
It may be concluded from the results that all the abundances were properly fitted even for
elements, such as the argon ions, where only one emission line was
observed for each ion. The extinction coefficient $c(H\beta)$ is
also considered in the analysis. Usually, the extinction cannot be
determined using collisionally excited lines due to their sensitivity
to temperature and density. In this case, however, we can see from
Fig. \ref{fig:test-case2-3} that $c\left(H\beta\right)$ was successfully
measured using only collisionally excited lines. There are two reasons
for that: first, two auroral lines, $\left[OIII\right]\lambda4363\AA$
and $\left[SIII\right]\lambda6312\AA$, are anchoring the electron
temperature. Second, this fitting includes emission lines from $\left[OIII\right]\lambda4363\AA$
to $\left[SIII\right]\lambda9531\AA$. This wide wavelength range
guarantees data points with varying sensitivity to the extinction,
which improves the sampling of $c\left(H\beta\right)$.

In the $3^{rd}$ test case, the simulation includes a prior $T_{high}$
for the high ionization region electron temperature. It should be
emphasised that in the previous test cases the ionised species had
the same assignment the low or high ionisation regions. The difference
now is that $T_{high}$ is fitted along the rest of the model parameters
instead of being calculated using eq. \ref{eq:TOIII-TSIII-relation}.
It can be appreciated in the right hand side of Fig. \ref{fig:test-case2-3}
that in this simulation there is a greater uncertainty in all the
parameter fittings. This is because the previous conditions are no
longer met. To begin with only one auroral line is anchoring each
electron temperature. Moreover, in this simulation only the $\left[OIII\right]$
and $\left[ArIV\right]$ emissions are located in the high ionisation
region. These lines cover a narrow wavelength range to sample the
parameter space which makes the $c\left(H\beta\right)$ sampling harder.
Finally, some small divergence on the traces for some parameters
during the simulation can be appreciated. We conclude that this test
case does not have enough data to fit the model parameters.

In the $4^{th}$ test case, the simulation runs the complete model:
three more dimensions $\left(y^{+},\,y^{2+},\tau\right)$ are included
reaching a total of fourteen. As all the hydrogen and helium lines
in the covered spectral region are included, the number of inputs
has also increased. Moreover, the hydrogen recombination lines provide
a firm anchor to the dust extinction since this parameter has the
greatest impact in their intensity. Fig. \ref{fig:test-case4} shows
that in this case the simulation has properly converged. Additionally,
it can be appreciated that all the parameter fittings are very close
to the true values tabulated in Table \ref{tab:testCasesResults}.

An important practice in Bayesian models is to check the impact of the priors design in the sampling process and results. The results from additional test cases for the complete model can be found in Table \ref{tab:extraTestCases}. In the $5^{th}$ test case the $T_{high}$ temperature prior has a uniform distribution which covers the complete temperature range: from $8000$ to $22000K$. In the $6^{th}$ test case both temperature priors have a Gaussian distribution but $T_{low}$ is centred at  $10000K$ while $T_{high}$ is centred at $20000K$. Both distributions have a standard deviation of $2500K$ which means that the true values at $15590K$ and $16000K$ have a relatively low probability. In the $7^{th}$ and  $8^{th}$ test cases we are assigning uniform priors for the optical depth and the electron density respectively. In the former case, the uniform priors lower and upper limits are 0 and 10 while in the latter the limits are 1 and 1000 $cm^{-3}$. Finally, in the $9^{th}$ test case the ionic helium abundance priors are defined by the relation: $y^{i+}=k_{y^{i+}}+Normal\left(\mu=0.0,\,\sigma=1.0\right)\cdot k_{y^{i+}}$, where $k_{y^{i+}} = 0.1$ and $k_{y^{i+}}=0.001$ for the $y^{+}$ and $y^{2+}$ abundances respectively. These priors represent a basic reparameterization: Distribution centred at zero with a certain offset. This prior design is not easy to read from a researcher point of view. Indeed, it may seem the obvious choice to assign a probability distribution which represents the physical parameter magnitude and behaviour as close as possible. Nevertheless, from the computational point of view, it is encouraged to use the same probability distributions centred at zero for all the model parameters and parameterizing the theoretical model instead. This is because the sampling process becomes more efficient. We can confirm in Table \ref{tab:extraTestCases} that the prior design is not affecting the results and the simulations are consistently converging at the same coordinate. These test cases represent an ideal scenario, where all the necessary inputs are available. In real observations where some emission lines may not be available the prior design impact needs to be checked again.

An additional graphical tool to establish a Monte Carlo simulation
convergence quality is a scatter plot matrix. This is shown in Fig.
\ref{fig:corner_plot_test-case4}, in which the parameter traces from
the complete test case are plotted against each other as surface distributions.
The red lines mark the true value location, which are within the algorithm
sampled region. It can be appreciated that there is some degeneracy
between the metals ionic abundance and electron temperature. This
degeneracy, however, is intrinsic for the physical model since for
the temperature and abundance range these parameters are correlated.
This degeneracy might be decreased by parameterizing the mathematical
parameter space, thus making it easier to sample. This would also
improve the speed of the model, which for the current implementation
is over 100 steps per second. However, the current algorithm convergence
quality and speed is good and the fittings presented display accuracy
and precision in the measurement. In the second phase, the algorithm
is tested on real observations and its convergence is compared to
the synthetic ones. This is done in the following section using the
data values from \citetalias{Fernandez2018b}.

\section{Results and discussion}

The initial implementations of this Bayesian algorithm in real spectra resulted in a slow sampling process or even failure to launch. This was caused by some emission lines which had a very narrow likelihood in the inference model. For example,
in this sample spectra the $\left[OIII\right]\lambda5007\AA$
and $H\alpha$ lines can display uncertainties below $0.5\%$ of their
integrated flux. Running synthetic test cases with similar likelihoods we learned that the algorithm found it hard to find a solution given the wide temperature and abundance priors provided. This uncertainty does not really represent the physical
processes intrinsic randomness. Instead, it quantifies the emission
line signal-to-noise ratio. Consequently, applying this uncertainty
in the emission line likelihood implies very small sampling steps for the model parameters. As most of the proposed parameter values are rejected the simulation runs very slowly. To deal with this issue a minimum $2\%$ uncertainty
was set on the emission lines likelihood. This is the value used in
the test cases, which provided stable solutions and measurements
with the expected uncertainty. 

\begin{figure*}
\includegraphics[width=1\textwidth]{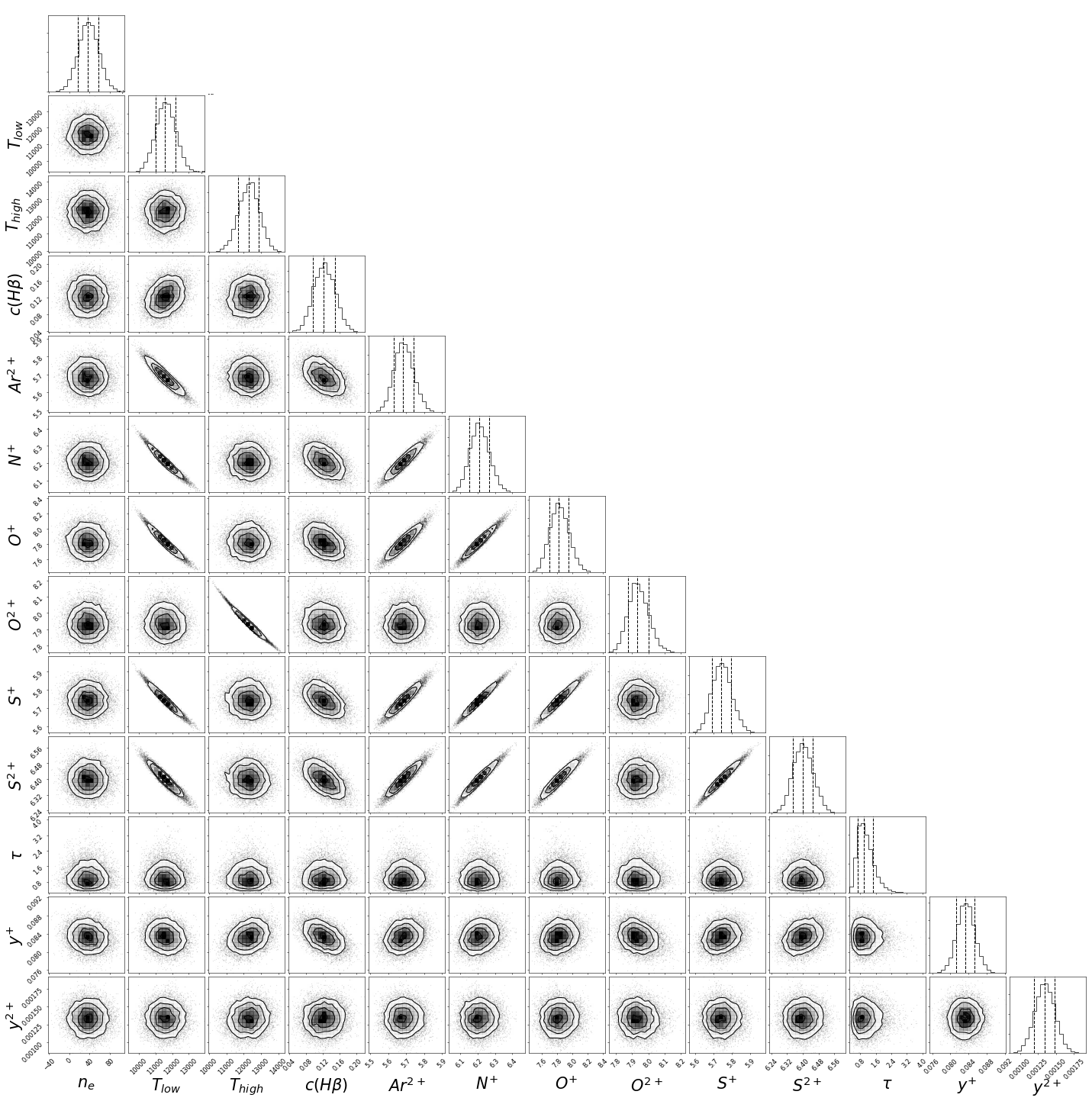}\caption{\label{fig:corner_plot_obj8}Scatter plot matrix for SHOC022. This
plot has been created using the corner library by \citet{Foreman-Mackey2016}.}
\end{figure*}
Fig.\ref{fig:corner_plot_obj8} corresponds to the fitting of SHOC022.
In most object fittings, the correlation between the ionic abundances
and the electron temperatures seem to be more pronounced than the
one displayed in Fig. \ref{fig:corner_plot_test-case4} for synthetic
inputs. Still, it can be established that the data is well represented
by the theoretical model and the correlation between parameters is
the one expected. An exception, however, is the optical depth $\tau$.
The surface plots in Fig.\ref{fig:corner_plot_obj8} for this parameter
imply a sampling limited by a mathematical boundary rather than a
true fitting. The reason behind these results can be found in the
available helium lines in the sample spectra. As it was explained
in \citetalias{Fernandez2018b}, due to a mismatch between our data
and the available Single Stellar Populations (SSP) synthesis models
wavelength range: only three $HeI$ lines include a correction for
the underlying stellar population ($HeI\lambda4471$, $5876$ and
$6678\AA$). These emission lines, however, are weakly affected
by fluorescence. As shown in \citet{Benjamin2002b}, the correction
$k_{\tau}$ in eq.\ref{eq:fluxFormula} for the $HeI6678\AA$
line accounts for less than $0.4\%$ even for the highest $\tau$
values. Therefore, it is very hard to fit $\tau$ in a parameter space,
where the other dimensions have a much larger impact in the final
line flux. Actually, the $\tau$ surface distributions in Fig.\ref{fig:corner_plot_obj8}
correspond to the input prior in Table \ref{tab:Priors-and-likelihood}:
a log-normal distribution with $\mu=0$ and $\sigma=0.4$. Therefore,
the optical depth cannot be measured from the provided observation
inputs. Nevertheless, the fluorescence excitation is still being taken
into account for this parameter and the simulation behaves as a standard
Monte Carlo algorithm for a user-declared parameter distribution.
This is an improvement from the helium abundance analysis in \citetalias{Fernandez2018b}.
The scatter plots for the complete HIIGs sample can be found in the
online support material.

\begin{figure}
\centering{}\includegraphics[width=0.8\columnwidth]{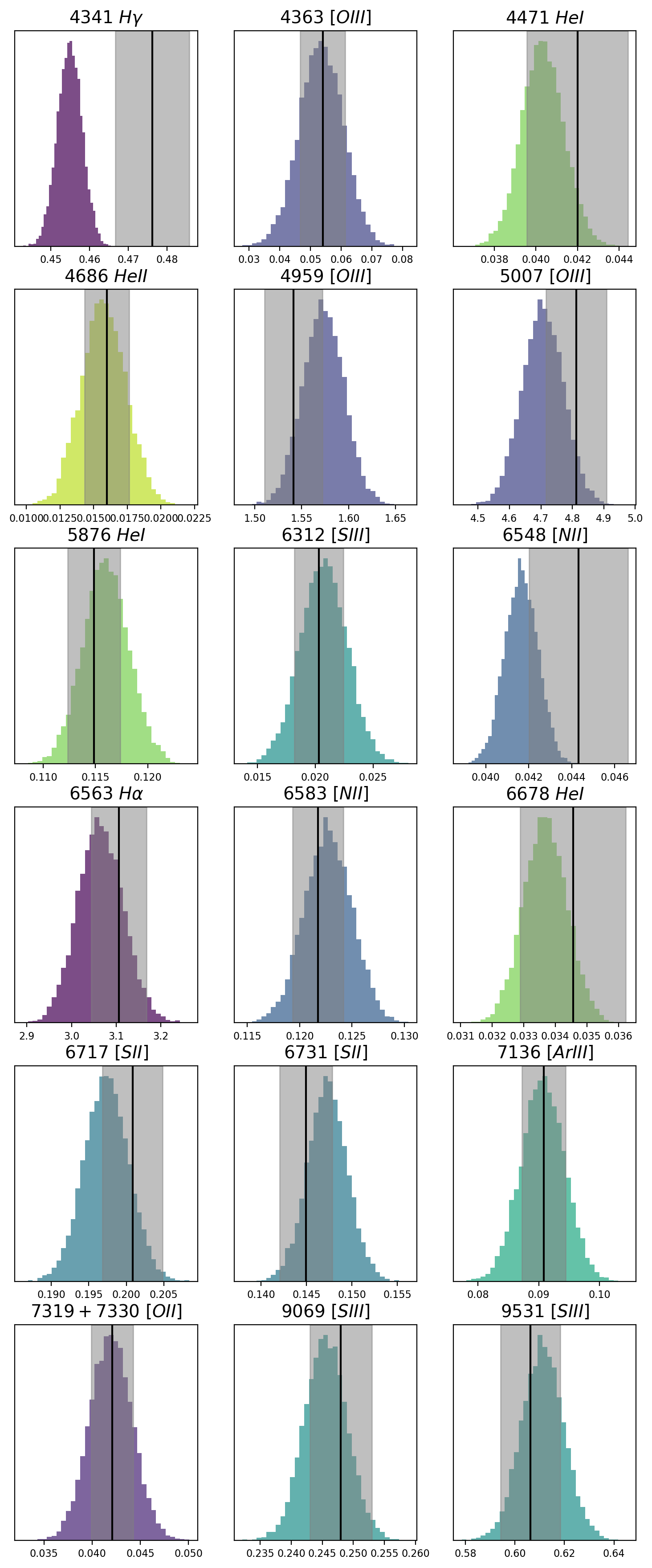}\caption{\label{fig:FluxPosteriors} Relative emission line flux posteriors
from the initial fitting of SHOC022. The output flux distributions
are plotted from blue to red wavelength. The distributions are colour
coded according to the ion producing each feature. The vertical black
lines represent the observed emission fluxes and the grey shaded area
their uncertainty. In colour in the electronic version.}
\end{figure}
Once the simulation convergence quality has been confirmed, the next
step consists in evaluating the fitting result. In a frequentist analysis
this can be done via the $\chi^{2}$ test. This is the approach followed
by \citet{Aver2015b} and references therein, where galaxies with
$\chi^{2}<5.5$ are excluded from the $Y_{P}$ regression. In the
Bayesian paradigm, there is not an universal estimator which can quantify
the fitting quality. However, in this particular analysis an efficient
evaluation can be accomplished by comparing the output flux distribution
with the input observed emission fluxes. This comparison can be found
in the tables included in the supporting online material. This information
can also be represented graphically and is shown in Fig.\ref{fig:FluxPosteriors}:
The fitted emission line flux distributions are plotted from bluest
to reddest wavelength. The distributions are colour coded according
to the ion responsible for each transition. Each plot cell includes
the observed line flux as a vertical line and its uncertainty as a
shaded area. These results correspond to the initial fitting of the
object. In general, it can be appreciated that the flux distributions
centre is very close to the observed value and their width is representative
of the measurement uncertainty. There are, however, some flux discrepancies
which require additional explanation:
\begin{itemize}
\item $H_{\gamma}$: The algorithm fails to fit this emission line. The
reason can be found in the instrumental setup: For most objects this
line lies right at the low wavelength edge. Consequently, the pixels
belonging to this line have a greater noise, which is not being propagated
by our current reduction pipeline. This was clearly observed during
the $c\left(H\beta\right)$ calculation in \citetalias{Fernandez2018b},
where this line did not match the extinction calculated from the comparison
between the $H\alpha$ to $H\beta$ flux ratio with the case B recombination
theoretical value. Consequently, this line have been excluded in most
objects fitting except for those with the higher redshift.
\item $HeI$ lines: not all helium lines are fitted with the same precision.
For example in SHOC022 the fitted flux for $HeI\lambda6678\AA$
is $6\%$ above the observed value while the $HeI\lambda4471\AA$
and $HeI\lambda5876\AA$ fluxes disagree only by a $0.5\%$ and
a $2\%$ respectively. In the particular case of the helium lines
this behaviour can be explained by the absorption from the underlying
stellar population. As it was shown in Fig. \ref{fig:Box-plot-with},
each line is affected differently by this effect. Additionally, in
the SSP synthesis applied in \citetalias{Fernandez2018b} the fitted
continuum does not include the uncertainty. Consequently, our current
methodology cannot quantify which helium line absorptions are better
fitted (as indeed was also the case in \citetalias{Fernandez2018b}).
The methodology presented in this paper, though, makes it much easier
to check which are the lines contributing more to the helium abundance
uncertainty.
\item $\left[OII\right]\lambda\lambda7319,7330\AA$ lines: In order
to include the recombination correction from \citet{Liu2001} our
algorithm takes as an input the integrated flux from both lines even
though our spectral resolution can separate the $\left[OII\right]$
doublet. 
\item $\left[NII\right]$ lines: Some of our objects show a broad $H\alpha$
component, that made difficult the deblending of the narrow $H\alpha$
and $[NII]$ lines as explained in \citetalias{Fernandez2018b}. The
observed ratio of the $[NII]$ doublet lines differ from the theoretical
value for some of the objects and large uncertainty was measured. Since the $\left[NII\right]\lambda6548\AA$ line is almost three times weaker than   $\left[NII\right]\lambda6583\AA$ and therefore harder to deblend, it was excluded in those fittings were a large mismatch was found in both lines fittings.
\item $\left[SIII\right]$ lines: Even though the $\left[SIII\right]\lambda6312\AA$
line is fitted with good precision, there is a small mismatch for
the infrared lines due to telluric contamination. As discussed in
\citetalias{Fernandez2018b}, even with the extra observations to
calibrate the sky features the ratio between the $\left[SIII\right]\lambda9069\AA$
and $\left[SIII\right]\lambda9531\AA$ lines is within $5\%$ of 
the expected value. Our theoretical model, however, does not include
the sky contribution in the $\left[SIII\right]$ computation, hence
the slight disagreement between the observed fluxes and the simulated
ones. Our analysis follows the same approach as in \citetalias{Fernandez2018b}:
both lines are included in the chemical analysis except in those cases
where the telluric calibration observations were not available. For
those objects, only the line less affected by the sky was used.
\end{itemize}
\begin{figure}
\includegraphics[width=1\columnwidth]{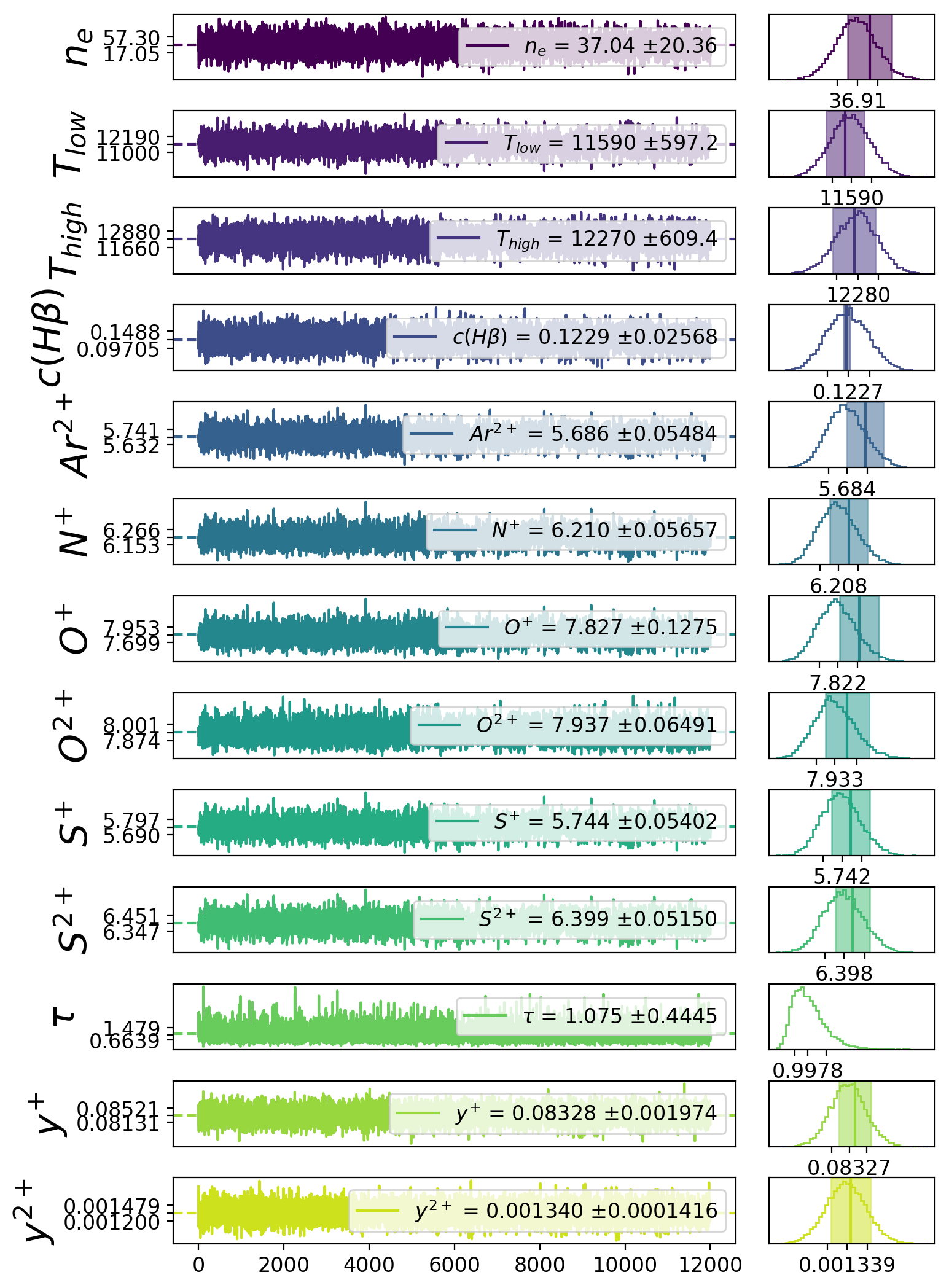}

\caption{\label{fig:SHOC022_posteriors}Posterior distributions for the fitting
of SHOC022. The vertical line in the right hand side histograms corresponds
to the parameter measurement published in \citetalias{Fernandez2018b}
using the traditional chemical analysis. In colour in the electronic version.}
\end{figure}
Once the discrepancies between the flux distributions generated by
the simulation and the observed line fluxes have been dealt with,
we may proceed with the next phase: evaluating the posterior distributions
for the model parameters. This provides an opportunity to compare
the results with those obtained classically. The results for SHOC022
are given in Table \ref{tab:SHOC022-HII-galaxy-table} in the appendix.
For the rest of the sample, the results are given online. In these
tables, column 1 corresponds to the parameter label, column 2 is the
measurement published in \citetalias{Fernandez2018b}, columns 3 and
4 are the trace mean and standard deviation, columns 5, 6 and 7 are
the $50^{th}$ (median), $16^{th}$ and $84^{th}$ percentiles and
column 8 is the percentage difference between the \citetalias{Fernandez2018b}
measurement and the present one. The graphical simulations presented
in section \ref{Model-test-cases} are also available for each
object. The one corresponding to the final fitting of SHOC022 can
be found in Fig. \ref{fig:SHOC022_posteriors} where the vertical
line in the histograms corresponds to the parameter measurement in
\citetalias{Fernandez2018b} and the shaded area represents its uncertainty.
It can be easily noticed that the $\tau$ posterior is actually the
prior log-normal as previously discussed. In general, however, most
parameters agree well, within the $5\%$, with those from \citetalias{Fernandez2018b}.
The following paragraphs discuss patterns encountered in most object
fittings:
\begin{itemize}
\item $T_{low}$ and $T_{high}$: We have shown in \citetalias{Fernandez2018b}
that, when one of the temperature diagnostic lines $\left[OIII\right]\lambda4363\AA$
or $\left[SIII\right]\lambda6312\AA$ had larger uncertainty
than the other, it was excluded from the analysis and the corresponding
temperature was calculated using eq. \ref{eq:TOIII-TSIII-relation}.
In order to duplicate the same procedure here, two actions were necessary.
First to replace the corresponding prior by eq. \ref{eq:TOIII-TSIII-relation}.
However, as discussed in section 4 second test case, once eq. \ref{eq:TOIII-TSIII-relation}
is added to the simulation both auroral lines, and their uncertainty,
dominate the temperature fitting. Therefore, as done in \citetalias{Fernandez2018b},
the second step consisted in excluding that line from the fitting.
\item $n_{e}\left[SII\right]$: In many objects there is a perfect match
between the new $n_{e}\left[SII\right]$ value and the one published
in \citetalias{Fernandez2018b}. This is expected since the density
prior is well constrained from the density measured via the $\left[SII\right]\lambda\lambda6717$,$6731\AA$
ratio. In some cases, including SHOC022 in Fig. \ref{fig:SHOC022_posteriors},
we obtain lower densities (e.g. $n_{e}\left[SII\right]=37\,cm^{-3}$
vs $n_{e}\left[SII\right]=50\,cm^{-3}$ in \citetalias{Fernandez2018b}
37\% smaller in this case). As it was described in \citetalias{Fernandez2018b} the reported density values in cases were $n_{e}\left[SII\right]<75\,cm^{-3}$  the density applied in the chemical analysis was $n_{e}\left[SII\right]=50\pm25\,cm^{-3}$. Therefore these results are consistent with those previously published.
\item $c\left(H\beta\right)$: We found that for some objects the extinction
measured via the current algorithm was higher and with a greater uncertainty
than those in \citetalias{Fernandez2018b}. In particular, MRK475
and PHL293,with very low extinction values in F2018 $\left(cH\beta\approx0.01\right)$
the new extinction coefficient measurements are one order of magnitude
larger. There are several reasons why this should be the case: to
start with, in the current analysis there are thirteen additional
parameters in the fitting. Secondly, while in the traditional methodology
$H\beta$ and $H\alpha$ dominate the $c\left(H\beta\right)$ calculation,
in the current algorithm all emission lines are contributing to its
measurement. In this study, the number of collisionally excited lines
is greater than the recombination ones. Consequently, they weight
heavily in the extinction determination. Nevertheless, as long as
the simulated line fluxes match the observed ones it can be concluded
that the new value quantifies the extinction adequately.
\item $Ar^{+3}$: This ion abundance measurement depends exclusively on
the $\left[ArIV\right]\lambda4741\AA$ observation. This is a
weak line even when compared with the helium ones. In the extreme
case of $IZw18$, the uncertainty for this line is above $20\%$ of
the observed flux. Due to these issues the sampler had problems to
fit the $Ar^{3+}$ abundance in this galaxy. Consequently, the prior
for this parameter was changed from the one in Table \ref{tab:Priors-and-likelihood}
to a normal distribution with $\mu=4.5$ and $\sigma=0.5$. This prior
provided a more informative sampling improving the fitting for this
parameter.
\end{itemize}

\subsection{Comparison between methods}

\begin{table}
\caption{\label{tab:SampleParameter_results}Sample fitting results. The values
displayed correspond to the mean and standard deviation for $n_{e}$,
the low and high ionization region $T_{e}$ and the extinction coefficient.}

\centering{}\begin{tabu}{lcccc}%
\hline%
&$n_{e}[SII]$&$T_{e}[SIII]$&$T_{e}[OIII]$&\\%
HII Galaxy&$(cm^{-3})$&$(K)$&$(K)$&$c(H\beta)$\\%
\hline%
FTDTR-1&42.1$\pm$20&14500$\pm$740&-&0.233$\pm$0.024\\%
IZw18&31.1$\pm$20&19000$\pm$1300&-&0.0567$\pm$0.019\\%
MRK36-A1&30.4$\pm$20&14900$\pm$610&14700$\pm$780&0.0743$\pm$0.022\\%
MRK36-A2&64.1$\pm$27&13000$\pm$290&15900$\pm$320&0.0516$\pm$0.019\\%
MRK475&36.5$\pm$19&14100$\pm$390&14600$\pm$320&0.0787$\pm$0.022\\%
FTDTR-2&39.5$\pm$22&-&13900$\pm$680&0.109$\pm$0.025\\%
IZw70&38.4$\pm$20&11200$\pm$500&-&0.0873$\pm$0.023\\%
MRK689&41.8$\pm$20&10300$\pm$390&-&0.266$\pm$0.028\\%
MRK67&70.5$\pm$31&12900$\pm$470&-&0.238$\pm$0.027\\%
FTDTR-3&45.7$\pm$19&13000$\pm$430&13000$\pm$320&0.101$\pm$0.023\\%
SHOC022&37.0$\pm$20&11600$\pm$600&12300$\pm$610&0.123$\pm$0.026\\%
FTDTR-4&45.0$\pm$20&12400$\pm$590&15100$\pm$310&0.186$\pm$0.024\\%
FTDTR-5&146$\pm$86&-&13600$\pm$1600&0.122$\pm$0.027\\%
FTDTR-6&45.9$\pm$20&12700$\pm$500&13000$\pm$380&0.181$\pm$0.025\\%
FTDTR-7&133$\pm$18&13300$\pm$350&13900$\pm$210&0.289$\pm$0.023\\%
MRK627&44.2$\pm$21&10500$\pm$470&11400$\pm$760&0.151$\pm$0.027\\%
PHL293B&113$\pm$27&15600$\pm$640&16600$\pm$230&0.225$\pm$0.027\\%
FTDTR-8&49.7$\pm$22&-&13100$\pm$1500&0.151$\pm$0.029\\%
SHOC263&233$\pm$60&-&11000$\pm$340&0.0685$\pm$0.021\\%
FTDTR-9&145$\pm$48&11800$\pm$870&11800$\pm$740&0.208$\pm$0.028\\%
FTDTR-10&582$\pm$80&-&16500$\pm$630&0.0850$\pm$0.023\\%
\hline%
\end{tabu}
\end{table}
\begin{table*}
\caption{\label{tab:HeON_ionic abundances}Helium, oxygen, and nitrogen ionic
abundances.}

\centering{}\begin{tabu}{lccccc}%
\hline%
HII Galaxy&$\nicefrac{He^{+}}{H^{+}}$&$\nicefrac{He^{2+}}{H^{+}}$&$12 + log\left(\nicefrac{O^{+}}{H^{+}}\right)$&$12 + log\left(\nicefrac{O^{2+}}{H^{+}}\right)$&$12 + log\left(\nicefrac{N^{+}}{H^{+}}\right)$\\%
\hline%
FTDTR-1&0.0975$\pm$0.0018&0.00117$\pm$0.00017&7.22$\pm$0.11&7.72$\pm$0.060&5.31$\pm$0.056\\%
IZw18&0.0752$\pm$0.0018&0.000527$\pm$0.00&6.59$\pm$0.11&7.00$\pm$0.065&4.79$\pm$0.057\\%
MRK36-A1&0.0767$\pm$0.0018&0.00180$\pm$0.00013&7.27$\pm$0.084&7.72$\pm$0.060&5.56$\pm$0.048\\%
MRK36-A2&0.0784$\pm$0.0015&0.00124$\pm$0.00&7.60$\pm$0.052&7.66$\pm$0.025&5.68$\pm$0.039\\%
MRK475&0.0821$\pm$0.0018&0.00171$\pm$0.00&7.28$\pm$0.062&7.79$\pm$0.028&5.68$\pm$0.042\\%
FTDTR-2&0.0750$\pm$0.0022&0.000963$\pm$0.00024&-&7.84$\pm$0.061&5.53$\pm$0.059\\%
IZw70&0.0939$\pm$0.0020&0.000570$\pm$0.00014&7.99$\pm$0.11&7.96$\pm$0.066&6.37$\pm$0.050\\%
MRK689&0.0795$\pm$0.0018&-&8.24$\pm$0.10&8.19$\pm$0.060&6.28$\pm$0.047\\%
MRK67&0.0836$\pm$0.0020&0.000964$\pm$0.00&7.63$\pm$0.083&7.95$\pm$0.048&5.84$\pm$0.049\\%
FTDTR-3&0.0833$\pm$0.0019&-&7.33$\pm$0.077&8.09$\pm$0.032&5.59$\pm$0.046\\%
SHOC022&0.0833$\pm$0.0020&0.00134$\pm$0.00014&7.83$\pm$0.13&7.94$\pm$0.065&6.21$\pm$0.057\\%
FTDTR-4&0.0865$\pm$0.0017&0.000744$\pm$0.00&7.73$\pm$0.11&7.82$\pm$0.025&6.00$\pm$0.050\\%
FTDTR-5&0.0764$\pm$0.0040&-&-&8.04$\pm$0.10&-\\%
FTDTR-6&0.0821$\pm$0.0017&0.000884$\pm$0.00&7.58$\pm$0.094&7.96$\pm$0.038&5.76$\pm$0.051\\%
FTDTR-7&0.0840$\pm$0.0015&0.000407$\pm$0.00&7.39$\pm$0.062&8.03$\pm$0.021&5.69$\pm$0.029\\%
MRK627&0.0869$\pm$0.0022&-&8.16$\pm$0.12&8.00$\pm$0.094&6.52$\pm$0.053\\%
PHL293B&0.0669$\pm$0.0018&0.00174$\pm$0.00017&6.89$\pm$0.086&7.62$\pm$0.019&-\\%
FTDTR-8&0.0737$\pm$0.0035&-&7.30$\pm$0.27&7.85$\pm$0.040&5.32$\pm$0.16\\%
SHOC263&0.0874$\pm$0.0042&-&8.04$\pm$0.093&8.01$\pm$0.047&6.59$\pm$0.045\\%
FTDTR-9&0.0876$\pm$0.0032&0.00160$\pm$0.00013&7.82$\pm$0.18&8.01$\pm$0.086&6.21$\pm$0.080\\%
FTDTR-10&0.0736$\pm$0.0031&0.000976$\pm$0.00014&6.98$\pm$0.095&7.53$\pm$0.046&5.84$\pm$0.048\\%
\hline%
\end{tabu}
\end{table*}
\begin{table*}
\caption{\label{tab:SA_ionic abundances}Sulphur and argon ionic abundances}

\centering{}\begin{tabu}{lccccc}%
\hline%
HII Galaxy&$12 + log\left(\nicefrac{S^{+}}{H^{+}}\right)$&$12 + log\left(\nicefrac{S^{2+}}{H^{+}}\right)$&$ICF\left(S^{3+}\right)$&$12 + log\left(\nicefrac{Ar^{2+}}{H^{+}}\right)$&$12 + log\left(\nicefrac{Ar^{3+}}{H^{+}}\right)$\\%
\hline%
FTDTR-1&5.15$\pm$0.046&5.83$\pm$0.046&1.93$\pm$0.10&5.13$\pm$0.052&5.14$\pm$0.074\\%
IZw18&4.70$\pm$0.048&5.15$\pm$0.047&1.19$\pm$0.041&4.59$\pm$0.057&3.85$\pm$0.12\\%
MRK36-A1&5.32$\pm$0.036&5.98$\pm$0.037&1.22$\pm$0.045&5.38$\pm$0.036&4.66$\pm$0.098\\%
MRK36-A2&5.43$\pm$0.024&6.16$\pm$0.026&1.17$\pm$0.018&5.49$\pm$0.024&4.64$\pm$0.049\\%
MRK475&5.38$\pm$0.027&6.13$\pm$0.030&1.15$\pm$0.021&5.51$\pm$0.028&4.57$\pm$0.068\\%
FTDTR-2&5.26$\pm$0.047&5.98$\pm$0.043&1.53$\pm$0.098&5.31$\pm$0.056&5.02$\pm$0.11\\%
IZw70&5.87$\pm$0.048&6.36$\pm$0.046&-&5.73$\pm$0.048&-\\%
MRK689&5.92$\pm$0.045&6.55$\pm$0.044&-&5.78$\pm$0.045&-\\%
MRK67&5.51$\pm$0.037&6.17$\pm$0.039&1.28$\pm$0.045&5.60$\pm$0.038&5.00$\pm$0.095\\%
FTDTR-3&5.21$\pm$0.034&6.14$\pm$0.035&1.93$\pm$0.086&5.51$\pm$0.034&5.48$\pm$0.033\\%
SHOC022&5.74$\pm$0.054&6.40$\pm$0.052&-&5.69$\pm$0.055&-\\%
FTDTR-4&5.61$\pm$0.048&6.22$\pm$0.046&1.15$\pm$0.024&5.58$\pm$0.048&4.67$\pm$0.066\\%
FTDTR-5&5.25$\pm$0.11&6.01$\pm$0.098&1.88$\pm$0.35&5.37$\pm$0.12&5.30$\pm$0.17\\%
FTDTR-6&5.41$\pm$0.040&6.27$\pm$0.040&1.23$\pm$0.039&5.60$\pm$0.040&4.88$\pm$0.076\\%
FTDTR-7&5.27$\pm$0.028&6.08$\pm$0.030&1.70$\pm$0.050&5.44$\pm$0.028&5.28$\pm$0.023\\%
MRK627&5.96$\pm$0.051&6.53$\pm$0.049&-&5.83$\pm$0.053&-\\%
PHL293B&4.96$\pm$0.036&5.73$\pm$0.039&1.60$\pm$0.10&5.14$\pm$0.037&4.91$\pm$0.082\\%
FTDTR-8&5.27$\pm$0.11&6.00$\pm$0.10&1.90$\pm$0.24&5.41$\pm$0.11&5.38$\pm$0.078\\%
SHOC263&5.96$\pm$0.036&6.56$\pm$0.043&-&5.81$\pm$0.036&-\\%
FTDTR-9&5.70$\pm$0.076&6.47$\pm$0.070&-&5.66$\pm$0.081&-\\%
FTDTR-10&5.39$\pm$0.038&6.07$\pm$0.034&-&5.26$\pm$0.043&-\\%
\hline%
\end{tabu}
\end{table*}
\begin{table*}
\caption{\label{tab:elemental_abundances}Element abundances and helium mass
fractions using either oxygen $Y_{\nicefrac{O}{H}}$ or sulphur $Y_{\nicefrac{S}{H}}$.}

\centering{}\begin{tabu}{lcccccc}%
\hline%
HII Galaxy&$\nicefrac{He}{H}$&$Y_{\left(\nicefrac{O}{H}\right)}$&$Y_{\left(\nicefrac{S}{H}\right)}$&$12 + log\left(\nicefrac{O}{H}\right)$&$12 + log\left(\nicefrac{N}{H}\right)$&$12 + log\left(\nicefrac{S}{H}\right)$\\%
\hline%
FTDTR-1&0.0987$\pm$0.0018&0.283$\pm$0.0037&0.283$\pm$0.0037&7.84$\pm$0.070&5.93$\pm$0.041&6.20$\pm$0.056\\%
IZw18&0.0757$\pm$0.0018&0.232$\pm$0.0042&0.232$\pm$0.0042&7.15$\pm$0.077&5.35$\pm$0.038&5.36$\pm$0.049\\%
MRK36-A1&0.0785$\pm$0.0018&0.239$\pm$0.0041&0.239$\pm$0.0041&7.85$\pm$0.050&6.15$\pm$0.063&6.15$\pm$0.035\\%
MRK36-A2&0.0796$\pm$0.0015&0.241$\pm$0.0035&0.242$\pm$0.0035&7.93$\pm$0.029&6.01$\pm$0.035&6.30$\pm$0.024\\%
MRK475&0.0838$\pm$0.0018&0.251$\pm$0.0040&0.251$\pm$0.0040&7.91$\pm$0.027&6.31$\pm$0.045&6.26$\pm$0.028\\%
FTDTR-2&0.0760$\pm$0.0022&-&0.233$\pm$0.0052&-&-&6.24$\pm$0.055\\%
IZw70&0.0944$\pm$0.0021&0.273$\pm$0.0044&0.274$\pm$0.0043&8.27$\pm$0.090&6.65$\pm$0.032&6.48$\pm$0.046\\%
MRK689&0.0795$\pm$0.0018&0.240$\pm$0.0042&0.241$\pm$0.0042&8.51$\pm$0.082&6.56$\pm$0.029&6.64$\pm$0.044\\%
MRK67&0.0846$\pm$0.0020&0.252$\pm$0.0043&0.253$\pm$0.0044&8.12$\pm$0.059&6.33$\pm$0.037&6.36$\pm$0.043\\%
FTDTR-3&0.0833$\pm$0.0019&0.249$\pm$0.0042&0.250$\pm$0.0042&8.16$\pm$0.031&6.41$\pm$0.053&6.47$\pm$0.027\\%
SHOC022&0.0846$\pm$0.0020&0.252$\pm$0.0044&0.253$\pm$0.0044&8.19$\pm$0.069&6.57$\pm$0.043&6.49$\pm$0.052\\%
FTDTR-4&0.0873$\pm$0.0017&0.258$\pm$0.0038&0.259$\pm$0.0038&8.08$\pm$0.053&6.36$\pm$0.024&6.37$\pm$0.042\\%
FTDTR-5&0.0764$\pm$0.0040&-&0.234$\pm$0.0093&-&-&6.35$\pm$0.091\\%
FTDTR-6&0.0830$\pm$0.0017&0.248$\pm$0.0039&0.249$\pm$0.0039&8.12$\pm$0.040&6.30$\pm$0.051&6.41$\pm$0.037\\%
FTDTR-7&0.0844$\pm$0.0015&0.252$\pm$0.0033&0.252$\pm$0.0034&8.12$\pm$0.021&6.42$\pm$0.031&6.37$\pm$0.023\\%
MRK627&0.0869$\pm$0.0022&0.257$\pm$0.0048&0.258$\pm$0.0048&8.39$\pm$0.085&6.75$\pm$0.045&6.63$\pm$0.049\\%
PHL293B&0.0687$\pm$0.0018&0.215$\pm$0.0044&0.215$\pm$0.0044&7.70$\pm$0.022&-&6.00$\pm$0.038\\%
FTDTR-8&0.0737$\pm$0.0035&0.227$\pm$0.0084&0.228$\pm$0.0084&7.97$\pm$0.082&6.00$\pm$0.16&6.35$\pm$0.071\\%
SHOC263&0.0874$\pm$0.0042&0.258$\pm$0.0092&0.259$\pm$0.0092&8.32$\pm$0.068&6.88$\pm$0.042&6.65$\pm$0.040\\%
FTDTR-9&0.0892$\pm$0.0032&0.262$\pm$0.0070&0.263$\pm$0.0070&8.23$\pm$0.095&6.63$\pm$0.069&6.54$\pm$0.071\\%
FTDTR-10&0.0746$\pm$0.0031&0.229$\pm$0.0072&0.230$\pm$0.0073&7.64$\pm$0.053&6.50$\pm$0.057&6.15$\pm$0.034\\%
\hline%
\end{tabu}
\end{table*}
 The fitting results for the complete sample are tabulated as follows:
the electron density $\left(n_{e}\left[SII\right]\right)$, the low
and
high ionisation temperatures and the logarithmic extinction coefficient
at $H\beta$ are shown in columns 2, 3, 4 and 5 in  Table \ref{tab:SampleParameter_results}. The optical depth,
defined as the mean value from the prior distribution, was found to be 
$\tau\approx1.05$
for most objects.
Table \ref{tab:HeON_ionic abundances} includes the $y^{+}$ and $y^{2+}$
abundances in columns 2 and 3, the oxygen abundances, columns 4 and 5 and nitrogen
$N^{+}$abundance in column 6. Table \ref{tab:SA_ionic abundances}
displays the sulfur ionic abundances in columns 2 and 3 and the argon
ionic abundances in columns 5 and 6. The $ICF\left(S^{3+}\right)$ is shown in Column 4 and was obtained 
following \citetalias{Fernandez2018b}, calculated via the
$S^{2+}$, $Ar^{2+}$, $Ar^{3+}$ abundances. Finally, Table \ref{tab:elemental_abundances}
displays the element abundances: column 2 corresponds to the helium
abundance while columns 3 and 4 correspond to the helium mass fractions
computed using the oxygen and sulfur abundances respectively. 
columns 5, 6 and 7 are the oxygen, nitrogen and sulfur abundances as $12+log\left(X\right)$.
We calculate the total abundances as in \citetalias{Fernandez2018b},
by adding the ionic abundances for each element using the traces as
in a standard Monte Carlo.

In general, these measurements agree very well with the ones published
in \citetalias{Fernandez2018b}. The differences between both methods
is below $3\%$ for most parameters except $n_{e}$ and $c\left(H\beta\right)$
as discussed above. A special case is the galaxy PHL293B. The comparison
between the simulation fluxes and the observed ones is as good as
for the rest of the sample. The parameter fittings, however, disagree
considerably from those in \citetalias{Fernandez2018b}. The measured
electron temperatures are $10\%$ and $5\%$ above the \citetalias{Fernandez2018b}
values for $T_{e}\left[SIII\right]$ and $T_{e}\left[OIII\right]$
respectively and the $c\left(H\beta\right)$ magnitude increased from
$0.01$ to $0.23$ once all the lines are taken in consideration.
\citet{Terlevich2014} presented WHT-ISIS and X-shooter spectra to
study the broad $\left(FWHM=1000\,\nicefrac{km}{s}\right)$ and very
broad $\left(FWHM=4000\,\nicefrac{km}{s}\right)$ components, as well
as the blue shifted and absorption components in $H\beta$ for this object. Reviewing
observations for this galaxy from 2005-2013 they concluded that this
uncommon emission must be caused by the young ionising cluster wind.
In a forthcoming paper we will try to adapt the emission model to
better fit this object's complex structure. 
FTDTR-10 also shows some  disagreement with the helium abundance previously
measured. In order to understand this discrepancy, it will be necessary
to repeat the chemical analysis using published spectra from other sources
such as the SDSS. For this paper, however, we exclude these galaxies
from the $Y_{P}-Z$ regressions. The simulation results for these
HIIGs are also available online.

The methodology applied in this paper presents the  advantage of being able to include (or
exclude) lines very easily and to quantify their impact on the target
abundances. This includes the chlorine and neon abundances not measured
in the current study. As discussed in section 4, this algorithm computes
ionic transition emissivity via the traditional approach: parameterised
equations as a function of the electron density and temperature. In
contrast, the treatment in \citetalias{Fernandez2018b} consisted
in a bilinear interpolation in a very fine grid generated with PyNeb for
each transition. For the hydrogen and metals considered both approaches
provide results which agree within 1\% for the complete temperature
and density domains considered. This is not the case for helium, that depends on
temperature and density in a remarkably more complex
manner. Some of the parametric equations considered for the \citet{Porter2012b}
recombination coefficients include those from \citet{Porter2007b},
\citet{Olive2004b} and \citet{Perez-Montero2017b}. None
of them guarantee a precision better than 1\% for the region of interest.
In order to improve the fitting, three temperature domains were considered:
the $8000\,K-12000\,K$, $12000\,K-15000K$ and $15000K-22000K$ ranges.
Even with this treatment, however, the discrepancy for some HeI emissivities goes up to $5\%$ at uneven $T_{e}$ and $n_{e}$ intervals. The
ideal solution consists of introducing a bilinear interpolation for
the emissivity calculation. Such an approach has been implemented
by \citet{Foreman-Mackey2019} in the exoplanet library. This will
be the approach to follow in the next algorithm iteration.

The simultaneous fitting of all the emission line fluxes provides
a more realistic analysis on how each one behaves. In general, the
helium lines were more likely to display a missmatch with their expected
flux values than those from the oxygen or sulfur ions. As explained
in section \ref{The-data} the flux absorption from
the underlying stellar population varies with wavelength.
This is a plausible cause for the disagreement in the $HeI$ lines.
A large discrepancy between the $HeI$ lines was observed in the two
starforming regions in $MRK36$. This galaxy is the only one in our
sample where two bursts were observed within the same slit. It is
possible that some cross talk in the spectra extraction is
contributing to the mismatch between the helium lines  measurement.
Additionally, the scheme applied in \citetalias{Fernandez2018b} to
fit the HIIGs continua does propagate the uncertainty introduced in
the $HeI$ lines intensity. Therefore, the current algorithm does
not quantify which recombination lines are more contaminated by the underlying 
stellar continua. There are additionally processes contributing
to the individual temperatures. For example, the $HeI\lambda5876\AA$
intensity may be contaminated by the $NaI$ $D_{1}$ and $D_{2}$
lines. Finally, a fluorescence contribution could not be properly
fitted by the available $HeI$ lines. These issues, however, are also
affecting the \citetalias{Fernandez2018b} and therefore the present $y$
and $y^{2+}$ measurements agree very well with those published in
\citetalias{Fernandez2018b}.

To the best of our knowledge, this is the first algorithm capable
of fitting both the recombination and collisionally excited line spectra
in a parameter space including the electron density and temperature.
There exist, however, many multi-dimensional chemical samplers based
on strong lines. An example of these libraries are HII-CHI-MISTRY
by \citet{Perez-Montero2014b}, IZI by \citet{Blanc2014b}, BOND by
\citet{ValeAsari2016b} or GAME by \citet{Ucci2018b}. Unlike the
algorithm presented here, these tools fit the input fluxes from photoionisation
model grids rather than a direct computation. These techniques provide
their own advantages. For example, they make possible the chemical
analysis of high redshift objects, they allow the modelling of more
complex temperature and density structures or can help to solve the
double value nature of some strong line diagnostics. There is a lot
to be learned from the techniques applied by these tools to explore
photoionisation grids. However, due to the high precision needed for
determining $Y_{P}$, strong line methods are not suitable for this
work. The strong line diagnostic strategy will be explored in future
developments of this algorithm. For example, instead of using the
fluorescence correction from \citet{Benjamin2002b}, the algorithm
could derive a better correction from photoionisation grids tailored
for each object. Currently, the closest analogues to our algorithm are
found in studies to determine  $Y_{P}$.

\begin{table}
\caption{\label{tab:Comparison-Aver}Comparison between the synthetic test
case published in \citet{Aver2015b} and one fitted with our
algorithm. The metal emission parameters have been excluded from this
comparison since they are not fitted by those authors methodology.
Similarly, the absorption on the hydrogen and helium lines are not
included since they are not fitted by our algorithm.}

\centering{}%
\begin{tabular}{cccc}
\hline 
Parameter & True value & Aver et al (2015) & Current work\tabularnewline
\hline 
$y^{+}$ & 0.085 & $0.0858\pm0.0027$ & $0.0851\pm0.0009$\tabularnewline
$y^{2+}$ & 0.00088 & \Xmark & $0.00088\pm0.00002$\tabularnewline
$n_{e}\left(He\right)$ & 500.0 & $473\pm67$ & $487\pm34$\tabularnewline
$\tau\left(HeI\right)$ & 1.0 & $0.78\pm0.31$ & $0.95\pm0.18$\tabularnewline
$T_{e}\left(He\right)$ & 16000.0 & $17320\pm1090$ & $16311\pm824$\tabularnewline
$c(H\beta)$ & 0.10 & $0.09\pm0.03$ & $0.106\pm0.015$\tabularnewline
$\xi\,\left(\times10^{4}\right)$ & 1.0 & $13_{+13}^{+19}$ & \Xmark\tabularnewline
\hline 
\end{tabular}
\end{table}
\citet{Izotov2004b} and \citet{Izotov2007b,Izotov2013b}
and references therein developed a  self-consistent helium abundance determination method.
This scheme consists in a Monte Carlo process chain to account for
the systematic effects on the $HeI$ lines. Izotov and colleagues vary stochastically
the $N_{e}\left(He^{+}\right)$, $T_{e}\left(He^{+}\right)$ and $\tau\left(3889\right)$
over a range of expected values to compute $y^{+}$. Afterwards, the
best solution for the helium abundance is fitted via a $\chi^{2}$test.
Each emission line is assigned a different underlying stellar absorption
and the helium lines are included in the $\chi^{2}$ minimisation. The
methodology introduced by \citet{Olive2001b} and \citet{Olive2004b}
and later enhanced in \citet{Aver2012a,Aver2013} and references therein,
also follows a frequentist paradigm applied to Izotov and collaborators
data. Their algorithm fits the nine dimensions space simultaneously
via a MCMC sampler. Both authors, however, emphasize how the $\chi^{2}$
fitting quality depends on the available $HeI$ lines sensitivity
to model parameters. For example, the data in \citet{Izotov2014b}
reaches the $HeI\lambda10830\AA$ line which, unlike most recombination
features, is very sensitive to the electron density. \citet{Aver2015b}
concluded after using these observations in their algorithm, that
the uncertainty in $Y_{P}$ decreased by more than $50\%$. 

As a final quality check, we may compare our results 
with the ones  by \citet{Aver2015b}. This can be easily done
by adding two additional dimensions to the model: $T_{e}\left(He\right)$
and $n_{e}\left(He\right)$. These are the electron temperature and
density contributing exclusively to the $HeI$ and $HeII$ emissivity.
In this $5^{th}$ test case, the $HeI\lambda3889$ \rm{and} $10830\AA$
lines have been included in the analysis in order to make the simulations
as similar as possible to each other. Table \ref{tab:Comparison-Aver}
compares the results from the synthetic test case in \citet{Aver2015b}
with the one solved using our HMC algorithm. This table does not include
the metals electron density and temperatures nor the ionic abundances
contributing to the collisionally excited lines since \citet{Aver2015b}
does not include them. Similarly, the absorption on hydrogen $\left(a_{H}\right)$
and helium $\left(a_{He}\right)$ are not tabulated since our algorithm
does not cover them. Once this is done, the number of parameters for
the recombination lines is very similar: the algorithm from Aver and
collaborators includes a correction for collisional excitation on
the hydrogen lines, which our current model does not. This correction
consists in a parametrisation by \citet{Anderson2000b} and \citet{Anderson2002b},
which depends on $T_{e}$ and the ratio of neutral to ionized hydrogen
atoms, $\xi$. On the other hand, our algorithm includes the $y^{2+}$
abundance which is measured from the $HeII4686\AA$ line. We
conclude from the results that both methods have similar accuracy
and precision. The precision on the helium temperature and density
is not as good as for the collisionally excited lines. This is expected
due to the low sensitivity of the recombination lines to these parameters.
Indeed this is the reason why the helium abundance is measured to
such a good accuracy despite the relatively large uncertainty on the
electron temperature. This is, though, not good enough for the primordial
helium abundance determination, where ideally one would like  to reach accuracies
better than $1\%.$ A more suitable prior for the helium temperature
could be designed by modelling the $T\left(He\right)$ priors from
the available $T_{e}\left[SIII\right]$ or $T_{e}\left[OIII\right]$
as it was done by \citet{Peimbert2002e}. Additionally, the width
in this temperature prior could also be added as a model dimension.
This would provide a 
quantification of the temperature
fluctuations, and thus, a direct comparison with the $Y_{P}$ regression
by \citet{Peimbert2017b}.

\subsection{$Y_{P}$ regression}

\begin{figure}
\begin{centering}
\includegraphics[width=1\columnwidth]{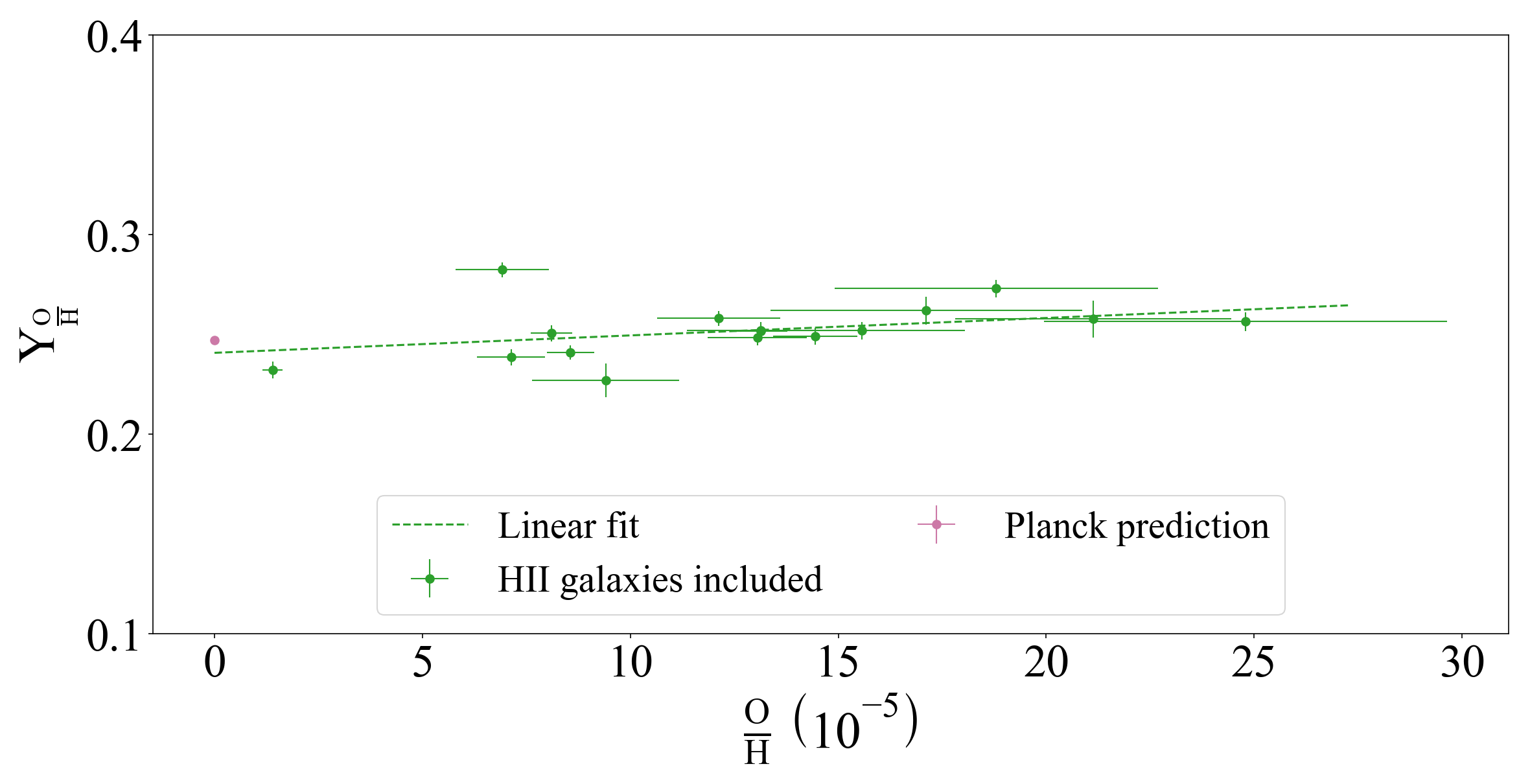}
\par\end{centering}
\begin{centering}
\includegraphics[width=1\columnwidth]{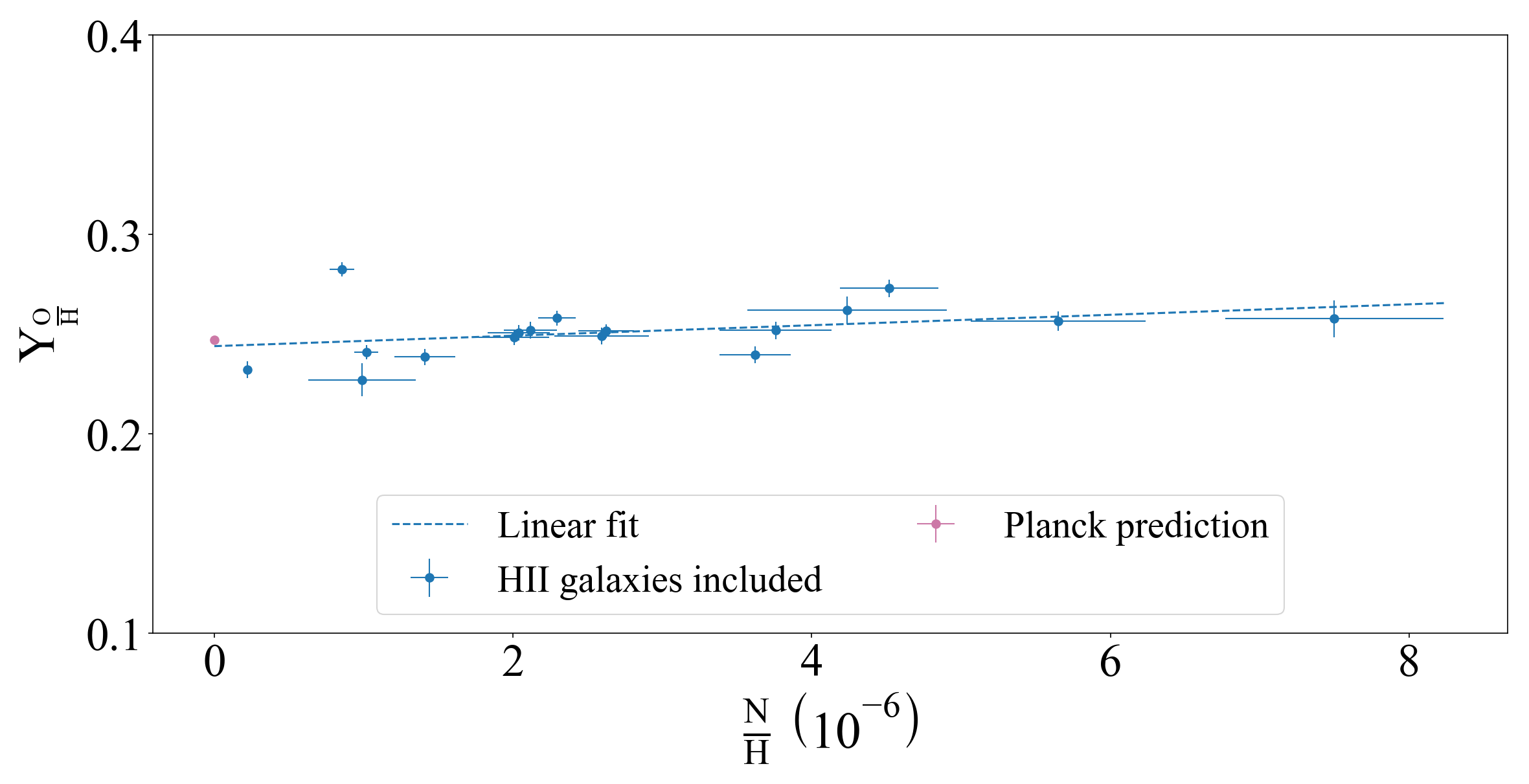}
\par\end{centering}
\begin{centering}
\includegraphics[width=1\columnwidth]{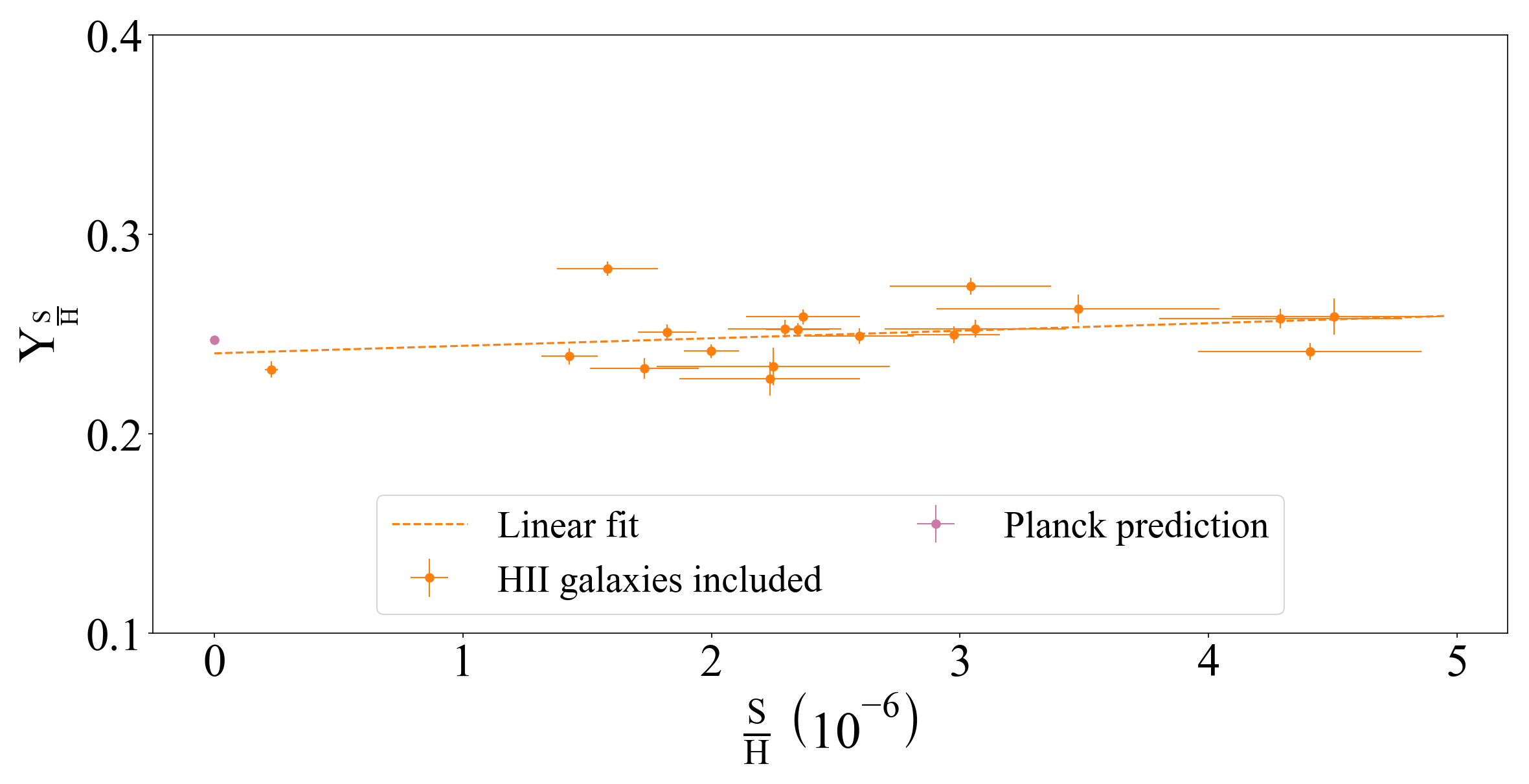}
\par\end{centering}
\caption{\label{fig:Primordial-linear-regressions}Primordial helium linear
regressions using oxygen, nitrogen and sulfur as metallicity tracer.}
\end{figure}
The helium mass fractions displayed in Table \ref{tab:elemental_abundances}
were calculated using the same procedure as in \citetalias{Fernandez2018b}. In summary,
for the oxygen and nitrogen $Y_{P}$ regressions, the $Y_{P,\,O}=f\left(\nicefrac{HeI}{H},\,\nicefrac{O}{H}\right)$
formulation from \citet{Peimbert1974} and \citet{Lequeux1979} was
applied. For the regression with sulfur the classical relation was
modified to:
\begin{equation}
Y_{P,\,S}=\frac{4\frac{He}{H}\left(1-20\cdot\frac{O}{S}\frac{S}{H}\right)}{1+4\frac{He}{H}}\label{eq:}
\end{equation}
where the $\nicefrac{O}{S}$ ratio is computed from the sulfur to
oxygen relation from the low metallicity BCDs (Blue Compact Dwarfs)
sample published by \citet{DorsJr.2016b}: $log\left(\nicefrac{S}{O}\right)=-1.53\pm0.03$.
As it was done in \citetalias{Fernandez2018b} the $Y_{P,\,O}$ value
is used in the $Y-N$ regression since the oxygen abundance is necessary
to compute the nitrogen one. A Monte Carlo algorithm described in
\citetalias{Fernandez2018b} was used to perform the $Y-O$, $Y-N$
and $Y-S$ regressions, as well as the multivariable regression using
the three metals. 

\textcolor{red}{}
\begin{table}
\textcolor{black}{\caption{\label{tab:Y_table}Primordial helium abundance determinations from
all the linear regression combinations and comparison with the literature.}
}
\begin{centering}
\textcolor{black}{\begin{tabu}{lcc}%
\hline%
Element regression&Value&Number of objects\\%
\hline%
$Y_{P,\,O}$&$0.241\pm0.004$&16\\%
\hline%
$Y_{P,\,N}$&$0.244\pm0.003$&17\\%
\hline%
$Y_{P,\,S}$&$0.240\pm0.004$&19\\%
\hline%
$Y_{P,\,O-N-S}$&$0.243\pm0.005$&16\\%
\hline%
\hline%
$Y_{P,\,O}^{1}$&$0.246\pm0.005$&18\\%
\hline%
$Y_{P,\,N}^{1}$&$0.251\pm0.005$&18\\%
\hline%
$Y_{P,\,S}^{1}$&$0.244\pm0.006$&21\\%
\hline%
$Y_{P,\,O-N-S}^{1}$&$0.245\pm0.007$&17\\%
\hline%
\hline%
$Y_{P,\,O}^{2}$&$0.2446\pm0.0029$&5\\%
\hline%
$Y_{P,\,O}^{3}$&$0.2449\pm0.0040$&15\\%
\hline%
$Y_{P,\,O}^{4}$&$0.2551\pm0.0022$&28\\%
\hline%
$Y_{P,\,Planck BBN}^{5}$&$0.24467\pm0.0002$&-\\%
\hline%
\end{tabu}}
\par\end{centering}
\textcolor{black}{{[}1{]} }\citetalias{Fernandez2018b}\textcolor{black}{{}
{[}2{]} \citet{Peimbert2016b} {[}3{]} \citet{Aver2015b} {[}4{]}
\citet{Izotov2014b} {[}5{]} \citet{PlanckCollaboration2018a} (This
value represents an upper limit from the four $\Lambda CDM$ parameter
configurations presented by the authors)}
\end{table}
Fig.\textcolor{black}{\ref{fig:Primordial-linear-regressions} shows
the regressions using oxygen, nitrogen and sulfur.}\textcolor{red}{{}
}The results from the linear fitting are listed in Table \ref{tab:Y_table}.
The first four rows correspond to the primordial helium abundance
computations using the chemical abundances from the Bayesian algorithm.
The following four rows correspond to the values published in \citetalias{Fernandez2018b}
using the traditional methodology. In the final rows in Table \ref{tab:Y_table}
we reproduce recent $Y_{P}$ determinations from the literature. It
can be concluded from this comparison that the $Y_{P}$ determinations
from both methodologies agree within their uncertainties. The largest
disagreement may be found in the $Y_{P,\,N}$ measurement. This could
be due to having excluded the $\left[NII\right]\lambda6548\AA$
line in some of the objects due to discrepancies with the observed values.
The current $Y_{P,\,N}$ result is closer to the oxygen and sulfur
regressions than the one given in \citetalias{Fernandez2018b}. Slightly lower values for
$Y_{P,\,O}$
and $Y_{P,\,S}$ can be appreciated also. This can be explained by the helium emissivity computation
as discussed previously. Nevertheless, the results match very well
those from the traditional determinations. Noticeably, the uncertainty
has decreased for all the $Y_{P}$ determinations. This is a remarkable
achievement, specially once it is considered how quickly and cleanly
the new methodology can complete a full direct method chemical analysis.
Since the new scheme fits all the ionic species simultaneously we
take the $Y_{P,\,O-N-S}$ determination as the preferred one:
\[
Y_{P,\,O-N-S}=0.243\pm0.005
\]
This result is consistent with the determinations from the Planck experiment in a framework provided by
a standard Big Bang cosmology.

\section{Conclusions}

The nebular abundances computed using the traditional direct method
are compared with those from a new Bayesian algorithm which simultaneously
fits a 14 parameters chemical model. The emission line fluxes belong
to the 21 HIIGs sample presented in \citetalias{Fernandez2018b} where
the recombination lines have been corrected by the underlying stellar
population. The model parameters are one electron density, two electron
temperatures (for the low and high ionization species), the logarithmic
extinction coefficient at $H\beta$, the optical depth for the $HeI$
transitions and nine ionic abundances: $Ar^{2+}$, $Ar^{3+}$, $y^{+}$,
$y^{2+}$, $O^{+}$, $O^{2+}$, $N^{+}$, $S^{+}$ and $S^{2+}$.
The sampling of this relatively big parameter space was successful
thanks to machine learning tools. The new HMC sampling shortens the
simulation length from several hours to a couple of minutes while
providing a stable convergence in contrast to the better known MCMC
samplers. The main conclusions from this analysis are:
\begin{itemize}
\item The direct method was adapted to a Bayesian paradigm by a careful
design for the model parameter priors. Uninformative (wide) probability
distributions were successfully applied for the electron temperatures,
the ionic abundances and the extinction coefficients. In contrast,
informative priors were necessary for the electron density and the
optical depth. For $n_{e}$ this was necessary due to the fact that
for these low density HIIGs sample (below the low density limit) 
none of the available collisionally
excited ratios are good diagnostic tools. On the other hand, The $HeI$
lines available in our data are weakly affected by the fluorescence
excitation. Consequently, the optical depth which parametrises it
could not be fitted in these observations. We
still took it into consideration while calculating the HeI fluxes
as if it were a standard Monte Carlo simulation.
\item A set of synthetic test cases with increasing number of model parameters
are presented to confirm the accuracy and precision of the sampler.
Once the algorithm is applied on real observations the behaviour and
correlation observed on the parameters remain constant. An issue,
however, was encountered during the fitting of real observations due
to the high precision  in the strong lines such as $\left[OIII\right]\lambda5007\AA$
and $H\alpha$. A minimum accuracy of $2\%$ was imposed in the observed
fluxes. The precision in the results is as good, though, as the one
expected from these high quality observations.
\item In order to test the success of the simulation, we propose
a graphical comparison between the input fluxes and the output flux
distributions from the fitting. We found in general a very good match
between the two. Some objects, however, displayed a sizeable mismatch
in the $H\gamma$ and $\left[NII\right]\lambda6548\AA$ fluxes
which can be explained by  technical constraints in the observations.
In those cases, the lines were excluded from the fitting. 
\item In general, we obtain a very good match with the abundances published
in \citetalias{Fernandez2018b}. Nevertheless, the new values for
the logarithmic extinction coefficient were higher for some objects
and with a consistent larger uncertainty. This can be  explained
by the difference in procedure: while in the traditional methodology
only $H\alpha$ and $H\beta$ end up contributing to the $c\left(H\beta\right)$
calculation in this new analysis all the lines are. In spite of this,
the abundances obtained are similar, as the extinction for these objects
is relatively small.
\item An additional synthetic test case was presented in order to reproduce
the test case in \citet{Aver2015b}. The $Te\left(He\right)$ and
$n_{e}\left(He\right)$ are added as additional parameters in the
$HeI$ and $HeII$ fluxes calculation. Despite the difference in mathematical
and computational schemes our algorithm can replicate the accuracy
and precision in the measurement of the parameters in common. It is
argued, however, that given the low dependency of the recombination
lines on electron temperature and density, the input prior design
may have a larger weight in the fitting results. In the near future we plan  to
explore new observations and theoretical models to better constrain
the $T_{e}\left(He\right)$ and $n_{e}\left(He\right)$ measurement.
\item The primordial helium abundance measurement from the new chemical
abundances is in good agreement with those in \citetalias{Fernandez2018b}.
The biggest disagreement was found for the $Y-N$ regression which
resulted in a lower $Y_{P,\,N}$ value obtained in the present work.
This is a consequence of the excluded $\left[NII\right]\lambda6548\AA$
lines due to a mismatch with the model fluxes. The present result, however,
is in good agreement with $Y_{P,\,O}$ and $Y_{P,\,S}$. Since this
technique fits all the ionic abundances simultaneously, we take the
multivariable linear regression as our chosen measurement giving $Y_{P,\,O-N-S}=0.243\pm0.005$.
This result is consistent with Standard Big Bang Nucleosynthesis.
\end{itemize}
We obtain in general lower uncertainties than we did with the classical
direct method. This new chemical analysis allows to increase the complexity
of the theoretical model. In future work, we will focus in including
third party data grids during the sampling process. This will improve
the emissivity computation for the helium lines which was one of the
challenges found in this paper. Finally, we hope to include the continua
computation in the fitting. This task has two purposes: a better quantification
of the stellar continuum absorption on the helium lines and a better
estimation of $T\left(He\right)$ via the nebular continuum jumps.
The algorithms developed in this work are not yet available in the
standard python distribution channels. However, all the scripts can
be found at the following github account until they are properly published
(https://github.com/Vital-Fernandez). 

\section*{Acknowledgements}

Vital Fern\'andez is indebted to Enrique P\'erez and Daniel Miralles
for their generous discussions on the nebular and stellar continua
determination, respectively. Similarly, he is grateful to Erick Aver
for his prompt replies on the physical and computational characteristics
of his helium abundance determination algorithm.

We are indebted to an anonymous referee whose detailed comments contributed to the improvement of the paper.

This work could not have been accomplished without the work and generosity
from the PyMC3 development team. V. F. also expresses his gratitude
to Dan Foreman-Mackey for his insight on how to use data science algorithms
in astronomical research.

We thank the Spanish allocation committee (CAT) for awarding observing
time and the cheerful technical support from the observatory personnel.
Vital Fern\'andez is grateful to the Mexican research Council (CONACYT)
for supporting this research through studentship 554031/300844 and Elena
and Roberto Terlevich acknowledge CONACYT for research
 grant CB-2008-103365. This work has been supported
by DGICYT grants AYA2013-47742-C4-3-P and AYA2016-79724-C4-1-P. Partial
financial support came also from project SELGIFS: PIRSES-GA- 2013-612701-SELGIFS.
Vital Fern\'andez and Elena and Roberto Terlevich are grateful to the
hospitality of the Departamento de F\'isica Te\'orica at the Universidad
Aut\'onoma de Madrid, Spain during a visit to advance with this study.

Funding for the creation and distribution of the SDSS Archive has
been provided by the Alfred P. Sloan Foundation, the Participating
Institutions, the National Aeronautics and Space Administration, the
National Science Foundation, the US Department of Energy, the Japanese
Monbukagakusho and the Max Planck Society. The SDSS Web site is
http://www.sdss.org.

This research has made use of the NASA/IPAC Extragalactic Database
(NED) which is operated by the Jet Propulsion Laboratory California
Institute of Technology, under contract with the National Aeronautics
and Space Administration.




\bibliographystyle{mnras}
\bibliography{Yp_BayesianModel_v0}



\appendix

\section{Extra material}

This section includes the tabulated results for the galaxy SHOC022 fitting presented in the discussion, as well as the atomic data references and their parametrisations.

\begin{table*}
\caption{\label{tab:SHOC022-HII-galaxy-table}SHOC022 fitting results.}
\begin{tabu}{lcccccccc}%
\hline%
Parameter&F2018 value&Mean&Standard deviation&Number of points&Median&$16^{th}$ percentile&$84^{th}$ percentile&Difference $\%$\\%
\hline%
$T_{low}$&11390&11590&597.2&12000&11590&11000&12190&1.703\\%
$n_{e}$&49.46&37.04&20.36&12000&36.91&17.05&57.30&-34.01\\%
$T_{high}$&12170&12270&609.4&12000&12280&11660&12880&0.9137\\%
$Ar^{2+}$&5.736&5.686&0.05484&12000&5.684&5.632&5.741&-0.9107\\%
$N^{+}$&6.238&6.210&0.05657&12000&6.208&6.153&6.266&-0.4898\\%
$O^{+}$&7.966&7.827&0.1275&12000&7.822&7.699&7.953&-1.836\\%
$O^{2+}$&7.972&7.937&0.06491&12000&7.933&7.874&8.001&-0.4868\\%
$S^{+}$&5.767&5.744&0.05402&12000&5.742&5.690&5.797&-0.4333\\%
$S^{2+}$&6.419&6.399&0.05150&12000&6.398&6.347&6.451&-0.3336\\%
$c(H\beta)$&0.1205&0.1229&0.02568&12000&0.1227&0.09705&0.1488&1.816\\%
$\tau$&None&1.075&0.4445&12000&0.9978&0.6639&1.479&None\\%
$y^{+}$&0.08388&0.08328&0.001974&12000&0.08327&0.08131&0.08521&-0.7379\\%
$y^{2+}$&0.001362&0.001340&0.0001416&12000&0.001339&0.001200&0.001479&-1.764\\%
$\frac{OI}{HI}$&8.270&8.192&0.06910&12000&8.188&8.124&8.261&-0.9948\\%
$\frac{NI}{HI}$&6.552&6.575&0.04346&12000&6.571&6.533&6.617&0.2797\\%
$\frac{SI}{HI}$&6.612&6.486&0.05178&12000&6.485&6.434&6.538&-1.954\\%
$\frac{HeI}{HI}$&None&0.08462&0.001979&12000&0.08460&0.08263&0.08656&None\\%
$Y_{O}$&0.2532&0.2521&0.004405&12000&0.2520&0.2477&0.2564&-0.4633\\%
$Y_{S}$&0.2542&0.2528&0.004418&12000&0.2528&0.2484&0.2572&-0.5568\\%
\hline%
\end{tabu}
\end{table*}

\begin{table*}
\caption{\label{tab:Parametrised-relations-and}Atomic data references for
the emission lines considered along with the parametrised relations.}

\centering{}%
\begin{tabular}{cccc}
\hline 
Ion & \multicolumn{2}{c}{Atomic data} & Emissivity parametrisation\tabularnewline
\hline 
\hline 
$H$ & \multicolumn{2}{c}{\citet{Storey1995b}} & $a+b\cdot log\left(T_{e}\right)+c\cdot log^{2}\left(T_{e}\right)$\tabularnewline
\hline 
$He$ & \multicolumn{2}{c}{\citet{Porter2013b}} & $\left(a+b\cdot n_{e}\right)\cdot log\left(t_{e}\right)-log\left(c+d\cdot n_{e}\right)$\tabularnewline
\hline 
$He^{+}$ & \multicolumn{2}{c}{\citet{Storey1995b}} & $a+b\cdot log\left(t_{e}\right)$\tabularnewline
\hline 
\hline 
Ion & Collision Strengths & Transition probabilities & Emissivity parametrisation\tabularnewline
\hline 
\hline 
$O^{+}$ & \citet{Pradhan2006b,Tayal2007b} & \citet{Zeippen1982b,Wiese1996b} & \multirow{2}{*}{$a+\frac{b}{T_{e}}\cdot log\left(T_{e}\right)+c\cdot log\left(T_{e}\right)+log\left(1+d\cdot n_{e}\right)$}\tabularnewline
\cline{1-3} 
$S^{+}$ & \citet{Tayal2010b} & \citet{Podobedova2009b} & \tabularnewline
\hline 
$O^{+2}$ & \citet{Aggarwal2000b} & \citet{Storey2000b,Wiese1996b} & \multirow{6}{*}{$a+b\cdot log\left(T_{e}\right)+c\cdot log\left(T_{e}\right)$}\tabularnewline
\cline{1-3} 
$N^{+}$ & \citet{Tayal2011b} & \citet{Wiese1996b,Galavis1997b} & \tabularnewline
\cline{1-3} 
$S^{+2}$ & \textcolor{black}{\citet{Hudson2012b}} & \citet{Podobedova2009b} & \tabularnewline
\cline{1-3} 
$S^{+3}$ & \citet{Tayal2000b} & \citet{Dufton1982b,Johnson1986b} & \tabularnewline
\cline{1-3} 
$Ar^{+2}$ & \citet{Galavis1995b} & \citet{Kaufman1986b,Galavis1995b} & \tabularnewline
\cline{1-3} 
$Ar^{+3}$ & \citet{Ramsbottom1997b} & \citet{Mendoza1982b} & \tabularnewline
\hline 
\end{tabular}
\end{table*}

\begin{table*}
\caption{\label{tab:Parametrised-coefficients}Emissivity parametrisation coefficients
for the relations given in Table \ref{tab:Parametrised-relations-and}. The Fitting precision for the HI, HeII and metal emissivities is better than $1\%$ for all the temperature and density ranges. In the HeI emissivities the precision decreases with up to a $5\%$ discrepancy in the $n_{e}$ and $T_{e}$ range of interest.}
\centering{}\textcolor{black}{\begin{tabu}{lccccc}%
\hline%
$Line$&a&b&c&d\\%
\hline%
$4341$ $H\gamma$&-0.846&0.230&-0.0251&-\\%
$4363$ $[OIII]$&4.97&-2.71&0.530&-\\%
$4471$ $HeI$&0.0153&0.000383&2.20&0.000041\\%
$4686$ $HeII$&1.09&-0.0630&-&-\\%
$4470$ $[ArIV]$&5.68&-1.21&0.736&-\\%
$4959$ $[OIII]$&5.23&-1.26&0.570&-\\%
$5007$ $[OIII]$&5.71&-1.26&0.570&-\\%
$5876$ $HeI$&0.000&0.000791&0.828&0.000052\\%
$6312$ $[SIII]$&5.25&-1.64&0.712&-\\%
$6548$ $[NII]$&5.12&-0.906&0.545&-\\%
$6563$ $H\alpha$&1.59&-0.492&0.0522&-\\%
$6583$ $[NII]$&5.59&-0.906&0.545&-\\%
$6678$ $HeI$&0.000637&0.000833&2.93&0.000109\\%
$6717$ $[SII]$&6.29&-0.924&0.423&0.00002\\%
$6731$ $[SII]$&6.14&-0.910&0.399&0.0001\\%
$7136$ $[ArIII]$&5.90&-0.820&0.506&-\\%
$7319 + 7330$ $[OII]$&4.84&-2.48&0.459&0.00021\\%
$9069$ $[SIII]$&5.47&-0.675&0.584&-\\%
$9531$ $[SIII]$&5.86&-0.675&0.584&-\\%
\hline%
\end{tabu}}
\end{table*}

\begin{table*}
\caption{\label{tab:extraTestCases}Fitting results for additional test cases with different priors design described in the text.}
\centering{}%
\begin{tabular}{ccccccc}
\hline 
Parameter & True value & Test 5 & Test 6 & Test 7 & Test 8 & Test 9\tabularnewline
\hline 
$T_{low}$ & 15590 & $15640\pm283$ & $15580\pm280$ & $15640\pm281$ & $15630\pm286$& $ 15630 \pm 282 $\tabularnewline
$n_{e}$ & 500 & $493\pm36$ & $493\pm36$ & $489\pm38$ & $498\pm38$& $ 497 \pm 38 $\tabularnewline
$S^{+}$ & 5.48 & $5.474\pm0.020$ & $5.478\pm0.019$ & $5.474\pm0.020$ & $5.476\pm0.020$& $ 5.476 \pm 0.020 $\tabularnewline
$S^{2+}$ & 6.36 & $6.354\pm0.023$ & $6.359\pm0.022$ & $6.354\pm0.022$ & $6.355\pm0.023$& $ 6.355 \pm 0.022 $\tabularnewline
$O^{+}$ & 7.80 & $7.793\pm0.037$ & $7.802\pm0.037$ & $7.795\pm0.037$ & $7.794\pm0.037$& $ 7.794 \pm 0.037 $\tabularnewline
$O^{2+}$ & 8.05 & $8.048\pm0.017$ & $8.047\pm0.017$ & $8.049\pm0.017$ & $8.049\pm0.017$& $ 8.049 \pm 0.017 $\tabularnewline
$Ar^{2+}$ & 5.72 & $5.715\pm0.020$ & $5.719\pm0.019$ & $5.715\pm0.019$ & $5.716\pm0.019$& $ 5.716 \pm 0.020 $\tabularnewline
$Ar^{3+}$ & 5.06 & $5.058\pm0.016$ & $5.057\pm0.016$ & $5.059\pm0.016$ & $5.059\pm0.015$& $ 5.059 \pm 0.015 $\tabularnewline
$N^{+}$ & 5.84 & $5.835\pm0.020$ & $5.839\pm0.019$ & $5.836\pm0.019$ & $5.836\pm0.019$& $ 5.836 \pm 0.019 $\tabularnewline
$c(H\beta)$ & 0.100 & $0.106\pm0.018$ & $0.104\pm0.018$ & $0.106\pm0.018$ & $0.105\pm0.018$& $ 0.105 \pm 0.018 $\tabularnewline
$T_{high}$ & 16000 & $16020\pm195$ & $16040\pm198$ & $16010\pm196$ & $16020\pm191$& $ 16020 \pm 193 $\tabularnewline
$y^{+}$ & 0.0850 & $0.085\pm0.001$ & $0.085\pm0.001$ & $0.085\pm0.001$ & $0.085\pm0.001$& $ 0.085 \pm 0.001 $\tabularnewline
$y^{2+}$ & 0.00088 & $0.00088\pm0.00001$ & $0.00088\pm0.00002$ & $0.00088\pm0.00002$ & $0.00088\pm0.00002$& $ 00088 \pm 00002$ \tabularnewline
$\tau$ & 1.0 & $0.987\pm0.228$ & $0.990\pm0.228$ & $1.030\pm0.281$ & $0.972\pm0.230$& $ 0.976 \pm 0.228 $\tabularnewline
\hline 
\end{tabular}
\end{table*}


\bsp	
\label{lastpage}
\end{document}